\documentclass[twocolumn,longbibliography,amsmath,amssymb,aps,prd,10pt,superscriptaddress,nofootinbib,noeprint,preprintnumbers]{revtex4-2}

\pdfoutput=1

\usepackage{graphicx, color}

\usepackage[letterspace=-10]{microtype} 

\usepackage{bm, amsmath, amsfonts}
\usepackage{multirow, tabularx, dcolumn}
\usepackage{mathtools} 
\usepackage{subdepth} 

\usepackage[utf8]{inputenc} 
\usepackage{hyperref}

\usepackage{xcolor}
\definecolor{jlab_red}{RGB}{192,39,45}
\definecolor{jlab_orange}{RGB}{249,102,0}
\definecolor{jlab_blue}{RGB}{47,122,121}
\definecolor{jlab_green}{RGB}{65,125,10}
\definecolor{jlab_grey}{RGB}{125,125,125}

\newcommand{\cm}{\ensuremath{\mathsf{cm}}}

\renewcommand{\star}{\ensuremath{\ast}}

\newcommand{\ccz}{\ensuremath{\chi_{c0}}}

\newcommand{\DD}{\ensuremath{{D\bar{D}}}}
\newcommand{\DsDs}{\ensuremath{D_s\bar{D}_s}}
\newcommand{\DDst}{\ensuremath{D\bar{D}^\star}}
\newcommand{\DstDst}{\ensuremath{D^\star\bar{D}^\star}}

\newcommand{\psiom}{\ensuremath{\psi\omega}}

\newcommand{\etce}{\ensuremath{\eta_c\eta}}
\newcommand{\etcep}{\ensuremath{\eta_c\eta^\prime}}

\newcommand{\tick}{\ensuremath{\checkmark}}

\hypersetup{%
pdftitle = {DDbar near threshold},
pdfsubject = {QCD},
pdfkeywords = {QCD, Hadron, Charm, Physics, Lattice, Meson, Scattering},
pdfauthor = {Hadron Spectrum Collaboration},
colorlinks = {true},
filecolor = {black},
linkcolor = {jlab_blue},
menucolor = {black},
citecolor = {jlab_green},
urlcolor = {jlab_green},
}{}

\begin{document}

\title{$D\bar{D}$ interactions are weak near threshold in QCD}

\author{David~J.~Wilson}\email{d.j.wilson@damtp.cam.ac.uk}
\affiliation{DAMTP, University of Cambridge, Centre for Mathematical Sciences, Wilberforce Road, Cambridge CB3 0WA, UK}
\author{Jozef~J.~Dudek}
\email{jjdudek@wm.edu}
\affiliation{Department of Physics, College of William and Mary, Williamsburg, VA 23187, USA}
\author{Robert~G.~Edwards}
\email{edwards@jlab.org}
\affiliation{\lsstyle Thomas Jefferson National Accelerator Facility, 12000 Jefferson Avenue, Newport News, VA 23606, USA}
\author{Christopher~E.~Thomas}
\email{c.e.thomas@damtp.cam.ac.uk}
\affiliation{DAMTP, University of Cambridge, Centre for Mathematical Sciences, Wilberforce Road, Cambridge CB3 0WA, UK}
\collaboration{for the Hadron Spectrum Collaboration}
\noaffiliation

\date{15 July 2026}

\begin{abstract}
\noindent
We study near-threshold $D\bar{D}$ scattering in $S$ and $D$-wave to determine whether or not resonances or bound states are present. Working in the approximation where charm-annihilation is forbidden, with two degenerate light-quark flavors and a heavier strange quark, isospin is a good quantum number, and the only other channel that is kinematically open is $\eta_c\eta$. 
Using lattice QCD we compute, as a function of varying light-quark mass, the $S$-matrix for coupled-channel ${\eta_c\eta - D\bar{D} }$ scattering and find only weakly interacting meson pairs. 
In contrast to several other studies, we find no evidence for any bound-state or resonance singularity in the energy region between the deeply-bound $\chi_{c0}(1P)$ state and the $D_s\bar{D}_s$ threshold.
\end{abstract}

\preprint{JLAB-THY-26-4597}

\maketitle


\section{Introduction}

Many apparently straightforward questions regarding what Quantum Chromodynamics (QCD) predicts for hadron-hadron scattering amplitudes remain unanswered. Such amplitudes may feature short-lived resonances above threshold or bound-states below threshold, potentially giving rise to enhancements in experimental production of the corresponding hadron-hadron pair.
Of particular interest to the current generation of experiments exploring the spectrum of excited hadrons are processes involving charm quarks. These are readily observed at LHCb, Belle-II, and BES-III, among others, where pairs of open-charm hadrons are formed in a variety of ways.
From both experimental and theoretical standpoints, the behavior of one the simplest systems, near-threshold $D\bar{D}$ scattering (in isospin-0) in $S$-wave, remains unclear.

Lying well below the $D\bar{D}$ threshold, the $\chi_{c0}(3415)$, which has only $c\bar{c}$ annihilation decays in QCD, is readily interpreted as something closely resembling a quark model $c\bar{c}(1P)$ state. Above this, the situation becomes much less clear. Quark model expectations are for $c\bar{c}(2P)$ states above 3900 MeV, which would suggest a single $\chi_{c0}'$ resonance well above $D\bar{D}$ threshold, with then a significant gap before the next quark model state.
In contrast to such a simple picture, experimentally there are claims for as many as four resonance states above the $\chi_{c0}(3415)$ and below 4 GeV, each appearing in either a single final state ($ J/\psi \, \omega$~\cite{Belle:2009and}, $D\bar{D}$~\cite{LHCb:2019lnr,LHCb:2020pxc}, $D_s \bar{D}_s$~\cite{LHCb:2022aki}) or a single production process, $e^+e^- \to J/\psi \, D\bar{D}$~\cite{Belle:2017egg}.

Theoretically, approaches which consider strong interactions between heavy mesons suggest attraction or even binding into \emph{meson-meson molecules} is possible~\cite{Molina:2009ct,Dong:2021juy}, usually via pseudoscalar or vector meson exchange. Significant effects are predicted near many meson-meson $S$-wave thresholds~\cite{Dong:2021juy}. Approaches based on a diquark-antidiquark color attraction, leading to compact \emph{tetraquarks}, suggest many states beyond the usual quark model counting, including with $J^{PC}$ quantum numbers other than those accessible to a meson-meson pair in $S$-wave~\cite{Maiani:2004vq,Lebed:2016yvr}. Approaches considering both meson-meson and quark-antiquark degrees of freedom suggest a wide range of possible mixing between these components~\cite{Eichten:2005ga,Barnes:2005pb,Ortega:2017qmg,Miyake:2025frsx-swwp}.

One difficulty in building an interpretation of the experimental spectrum, is that none of the currently explored experimental processes that ultimately produce $D\bar{D}$ pairs are completely straightforward to analyse. Some have a complex progenitor such as a heavy $B$-meson decaying through a three-hadron process, where interactions with the third hadron can have a non-trivial impact. Another possibility is production via two-photon fusion~\cite{Belle:2005rte,BaBar:2010jfn}, and here the Born diagram with a $t$-channel $D$ exchange has the possibility to significantly distort the energy dependence near $\DD$ threshold relative to the $\DD$ elastic scattering amplitude~\cite{Wang:2020elp,Deineka:2021aeu}. Such distortion is well-known in the $\gamma\gamma\to\pi\pi$ and $\gamma\gamma\to K\bar{K}$ processes, which bear little resemblance to the $S$-wave $s$-channel $\pi\pi\to\pi\pi$ and $\pi \pi \to K\bar{K}$ processes~\cite{Hoferichter:2011wk,Dai:2014zta,Danilkin:2018qfn}.

\medskip
In the past 20 years, many of the new resonances claimed experimentally have been observed close to \mbox{$S$-wave} hadron-hadron thresholds. In order to answer questions about whether or not this is accidental, or a hint at an important poorly-understood dynamical aspect of QCD, we must obtain clean first-principles predictions from the underlying theory of QCD.

Lattice QCD offers the opportunity to do just this, making only controlled approximations to QCD, such as the use of a lattice spacing cutoff which can ultimately be removed. Use of a finite spatial volume means the theory has a discrete spectrum, and this spectrum can be used to constrain hadron-hadron scattering amplitudes~\cite{Luscher:1990ux,Kim:2005gf,Briceno:2014oea}.  The theory can be studied as a function of quark mass, exploring the sensitivity of observables to these parameters whose particular values are external to QCD.

A detailed lattice QCD study using light quarks with mass such that $m_\pi \sim 391$ MeV was reported in Refs.~\cite{Wilson:2023hzu,Wilson:2023anv}. Analysing finite-volume spectra up to the $J/\psi\, \phi$ threshold, coupled-channel amplitudes describing all relevant kinematically--open channels were determined, and in these no near-threshold resonances or bound states were found. The only $S$-wave singularity required was found to be a quark--model--like state appearing above the $D_s\bar{D}_s$ threshold\footnote{The PDG~\cite{ParticleDataGroup:2024cfk} currently combines the narrow resonances observed experimentally in $J/\psi\, \omega$, $D\bar{D}$ and $D_s \bar{D}_s$ into a single entry, the $\chi_{c0}(3915)$. The only additional experimental scalar candidate in this energy region is the broad $\chi_{c0}(3860)$ claimed in ${e^+e^- \to J/\psi \, D\bar{D}}$, but not required in descriptions of $B \to K \, D\bar{D}$.}.

The result of Refs.~\cite{Wilson:2023hzu,Wilson:2023anv} was in stark contrast to an earlier lattice QCD calculation, presented in Ref.~\cite{Prelovsek:2020eiw}, using 280 MeV pions, which concluded that all of the following were required in order to describe its finite-volume spectrum: a stable bound-state 4 MeV below $D\bar{D}$ threshold, an extremely narrow resonance on top of the $D_s\bar{D}_s$ threshold, and another broader resonance above $D\bar{D}$ threshold\footnote{While the analysis of Ref.~\cite{Prelovsek:2020eiw} used piecewise amplitudes whose analyticity could be questioned, reanalysis of the energy levels of Ref.~\cite{Prelovsek:2020eiw} using more consistent amplitudes led to a comparable state content~\cite{Shi:2024llv}, although the quality of fit of the finite-volume spectrum description is poor.}.  

While unlikely, it remained possible that these two calculations were in fact compatible, through the $D\bar{D}$ scattering process having an unexpectedly strong dependence upon the light-quark mass. 
It is to this possibility that we turn in this paper, by considering the techniques of Refs.~\cite{Wilson:2023hzu,Wilson:2023anv} at three lighter pion masses, \mbox{$m_\pi \sim 239,\, 283$ and 330~MeV}. We find that, contrary to the observations presented in Ref.~\cite{Prelovsek:2020eiw}, there appear to be no significant finite-volume energy shifts in levels near to the $D\bar{D}$ threshold, and that the resulting scattering amplitudes feature no nearby singularities corresponding to bound-states or resonances\footnote{While in this calculation we do not determine amplitudes to higher energies (e.g. above the $D_s \bar{D}_s$ threshold), we expect the picture observed in Refs.~\cite{Wilson:2023hzu,Wilson:2023anv} to continue to hold, with a single narrow resonance comparable to the $\chi_{c0}(3915)$.}.

\section{Lattice QCD setup}\label{sec:lqcd}

Our calculational approach follows that of Ref.~\cite{Wilson:2023anv}, where details can be found. In this paper we report on the use of three different light quark mass choices using lattices whose properties are summarized in Table~\ref{tab:lat}. These are anisotropic lattices with $N_f=2+1$ flavours of dynamical quarks, with the action described in Refs.~\cite{Edwards:2008ja,Lin:2008pr}. The valence charm quarks use the same Wilson-clover action as the light and strange quarks. 
The lattice scale is determined by setting the computed $\Omega$ baryon mass to its physical value, such that each lattice has spacing close to $a_t \sim 0.033\, \mathrm{fm}$ temporally and $a_s \sim 0.12\, \mathrm{fm}$ spatially. The charm quark is tuned by adjusting the bare charm quark mass parameter on each lattice such that the $\eta_c$ mass is close to its physical value~\cite{Liu:2012ze,Edwards:2012fx}.

A large basis of operators includes single-meson-like constructions formed from gamma matrices and up to three gauge covariant derivatives~\cite{Dudek:2010wm} and meson-meson-like constructions formed from product pairs of optimized single-meson-like constructions~\cite{Thomas:2011rh,Dudek:2012gj}. Quark fields are smeared with \emph{distillation~}\cite{Peardon:2009gh} with the number of Laplacian eigenvectors given in Table~\ref{tab:lat}. $c\bar{c}$ annihilation is forbidden by excluding Wick contractions in which it occurs, but light and strange annihilation is included without further approximation. 
Discrete spectra follow from diagonalizing a computed matrix of correlation functions, solving a generalized eigenvalue problem~\cite{Michael:1985ne,Luscher:1990ck}. Fits to the time-dependence of the eigenvalues are performed over various time-windows, and the energies used correspond to a weighted average over many such fits (see Appendix E of Ref.~\cite{Ortega-Gama:2024rqx}).

Our basis of operators includes constructions resembling $\eta_c\eta$, $\DD$, $\DsDs$, $\etcep$, and $\psiom$. The thresholds for the last three of these lie above the $\etce-\DD$ scattering region that is our main focus in this study. Including higher-lying operators helps to give us confidence in the extracted spectrum. In practice we find that excluding $\etcep$ operators altogether leaves the spectrum below $D_s \bar{D}_s$ threshold unchanged\footnote{As in Ref.~\cite{Wilson:2023anv}, correlators featuring $\etcep$ operators are found to be noisy and when included, the indications are for an approximate decoupling of the related channel.}. Stable hadron masses and hadron-hadron channel thresholds for the three different light quark masses are presented in Table~\ref{tab:masses}. We also list some three and four-hadron channels that can in principle contribute, but in each case we expect them to be highly suppressed either by large required values of $\ell$ or by their high-lying thresholds\footnote{We do not include any operators resembling three-hadron channels, such as $\ccz\pi\pi$ or $\DD\pi$.}.

The relevant symmetries of the periodic cubic volume are reflected in the use of lattice \emph{irreps}, with a detailed description for the current case given in Ref.~\cite{Wilson:2023anv}. In this study we compute only a limited set of irreps, $[000]A_1^+$, $[001]A_1$ and $[000]E^+$, that we find is able to provide enough constraint to determine $S$-wave and $D$-wave scattering in the energy region of interest. Due to the modest energy range considered, the basis of operators used (Appendix~\ref{app:ops_lists}) is a relatively small subset of those used in Ref.~\cite{Wilson:2023anv}, which considered amplitudes to a much higher energy.

\begin{table*}[!ht]
\begin{tabular}{c|rlccc|lll|c}
$-a_tm_\ell$ & $(L/a_s)^3$ & $\times \; T/a_t$ & $N_{\mathrm{cfgs}}$ & $N_\mathrm{vecs}$ & $N_\mathrm{t_{src}}$  
              & \multicolumn{1}{c}{$a_t m_\pi$} &  \multicolumn{1}{c}{$a_t m_\Omega$} & \multicolumn{1}{c|}{$\xi_\pi$} & $m_\pi$/MeV\\[0.2ex]
\hline
\hline
0.0860 & $\Bigl.\Bigr.$ $32^3$ &$\times$ 256 & 485 & 384 & 2--4  & 0.03928(18) & 0.2751(6)  & 3.453(6)     & 239  \\
0.0856 & $\Bigl.\Bigr.$ \{$24^3$, $32^3$\} &$\times$ 256 & \{400, 481\} & \{160, 256\} & 4--8     & 0.04720(11) & 0.2793(8)  & 3.457(6)     & 283  \\
0.0850 & $\Bigl.\Bigr.$ $24^3$ &$\times$ 256 & 481 & 160 & 4--8  & 0.05635(14)     & 0.2857(8)  & 3.456(9)     & 330  \\
\hline
\multirow{2}{*}{\it 0.0840} & $\Bigl.\Bigr.$ {\{$16^3$, $24^3$\}} & $\times$ 128    &  { \{478, 553\} }           & { \{64, 160\}} 
                                  & \multirow{2}{*}{\it 2--8}  & \multirow{2}{*}{\it 0.06906(13)} & \multirow{2}{*}{\it 0.2951(22)} & \multirow{2}{*}{\it 3.444(6)}     & \multirow{2}{*}{\it 391}   \\
           & $20^3$ & $\times$ 256 & 288 & 128 & & & & \\
\hline
\end{tabular}
\caption{
Details of the $N_f=2+1$ anisoptropic lattices used in this study corresponding to the actions described in Refs.~\cite{Edwards:2008ja,Lin:2008pr}. Results on the lattices in the bottom row were presented previously in Refs.~\cite{Wilson:2023hzu,Wilson:2023anv}. $N_{\mathrm{cfgs}}$ is the number of gauge configurations used, $N_\mathrm{vecs}$ is the number of vectors used in the distillation approach, and $N_\mathrm{t_{src}}$ are the number of time sources averaged in computing the correlation matrix.
}
\label{tab:lat}
\end{table*}

\begin{table*}[!ht]
\renewcommand{\arraystretch}{1.14}

\begin{tabular}{clll}
\multicolumn{1}{c}{} & \multicolumn{1}{l}{$239$ MeV} & \multicolumn{1}{l}{$283$ MeV} & \multicolumn{1}{l}{$330$ MeV} \\[1ex]
\hline\\[-1.5ex]
$\quad\pi$             & 0.03928(18)  & 0.04720(11)  & 0.05635(14)\\
$\quad K$               & 0.08344(7)   & 0.08659(14)  & 0.09027(15)\\
$\quad\eta$            & 0.09299(56)  & 0.09427(70)  & 0.09790(100)\\
$\quad{\eta^\prime}$   & 0.15666(200) & 0.16052(140) & 0.16286(156)\\
$\quad\omega$          & 0.13616(107) & 0.14081(61) &  0.14710(86) \\[1ex]
\hline\\[-1.5ex]
$\quad{D}$        & 0.30923(11) & 0.31406(11) & 0.32255(11)\\
$\quad{D_s}$      & 0.32356(12) & 0.32785(6)  & 0.33548(8)\\
$\quad{D^\ast}$   & 0.33058(24) & 0.33572(16) & 0.34408(16)\\
$\quad{D_s^\ast}$ & 0.34448(15) & 0.34877(8)  & 0.35641(12)\\[1ex]
\hline\\[-1.5ex]
$\quad{\eta_c}$   &  0.49097(4) & 0.49773(3)  & 0.50973(4)\\
$\quad{J/\psi}$   &  0.50553(5) & 0.51218(3)  & 0.52391(5)\\
\end{tabular}
$\quad$
\renewcommand{\arraystretch}{1.24}
\begin{tabular}{c lll}
\multicolumn{1}{c}{}  & \multicolumn{1}{l}{$239$ MeV} & \multicolumn{1}{l}{$283$ MeV} & \multicolumn{1}{l}{$330$ MeV} \\[1ex]
\hline\\[-1.5ex]
$\quad\etce$    & 0.5840(6)   & 0.5920(7)  & 0.6076(10) \\
$\quad\DD$      & 0.6185(2)   & 0.6281(2)  & 0.6451(2)  \\
$\quad\DDst$    & 0.6398(3)   & 0.6498(2)  & 0.6666(2)  \\
$\quad\psiom$   & 0.6417(11)  & 0.6530(6)  & 0.6710(9)  \\
$\quad\DsDs$    & 0.6471(2)   & 0.6557(1)  & 0.6710(1)  \\
$\quad\etcep$   & 0.6476(20)  & 0.6583(14) & 0.6726(16) \\
$\quad\DstDst$  & 0.6612(3)   & 0.6714(2)  & 0.6882(2)  \\[2.0ex]
$\quad\eta_c\pi\pi$     & 0.5695(3)   & 0.5921(2)  & 0.6224(2) \\
$\quad J/\psi\, \pi\pi\pi$ & 0.6234(3)   & 0.6538(2)  & 0.6930(2) \\
$\quad\ccz\pi\pi$       & 0.6463(3)   & 0.6697(4)  & 0.7015(3) \\
$\quad\DD\pi$           & 0.6577(2)   & 0.6753(2)  & 0.7015(2)  \\
\end{tabular}
\caption{Left: Stable hadron masses in temporal lattice units, $a_t m$. Right: Two, three and four hadron thresholds in temporal lattice units.}
\label{tab:masses}
\end{table*}

Independent determination of the energy of stable hadrons at various values of three-momentum leads to a measure of the dispersion relation from which an estimate of the lattice anisotropy, $\xi = a_s/a_t$, can be extracted. While each hadron is found to obey the relativistic dispersion relation to an excellent approximation, the values of the anisotropy extracted exhibit a modest scatter across different hadrons. We choose to propagate a systematic uncertainty due to that variation through the determination of the scattering amplitudes. Since we focus in this paper mainly on \emph{near-threshold} $D\bar{D}$ scattering, in practice we find relatively little sensitivity to this conservative approach. 
Table~\ref{tab:xi} presents the determined $\xi$ values for relevant stable hadrons, and the value for each pion mass (with inflated error) to be used later.

\begin{table}

\renewcommand{\arraystretch}{1.24}

\begin{tabular}{llll}
\multicolumn{1}{c}{}  & \multicolumn{1}{l}{$239$ MeV\;\;\;} & \multicolumn{1}{l}{$283$ MeV\;\;\;} & \multicolumn{1}{l}{$330$ MeV} \\[1ex]
\hline\\[-1.5ex]
$\xi_{\pi}$    & 3.453(6)  & 3.457(6)  & 3.456(9)  \\
$\xi_{\eta}$   & 3.468(20)  & 3.408(10) & 3.438(9)  \\
$\xi_{D}$      & 3.443(7)  & 3.432(4)  & 3.448(6)  \\
$\xi_{\eta_c}$ & 3.456(4)  & 3.457(2)  & 3.477(2)  \\[1ex]
\hline\\[-2ex]
$\xi$          & 3.453(17) & 3.457(25) & 3.456(14) \\
\end{tabular}

\caption{Anisotropies determined for the $\pi$, $\eta$, $D$ and $\eta_c$ mesons. Values for $m_\pi\sim 239$~MeV taken from Refs.~\cite{Wilson:2015dqa} and~\cite{Cheung:2016bym}. The final row indicates the estimate used in the scattering analyses in this study, whose central value comes from the pion determination with an error inflated to account for the variation above.}
\label{tab:xi}
\end{table}

We also add an additional uncertainty (${a_t\, \delta E_\mathrm{syst}=0.00050}$) to every energy level determined on the $m_\pi\sim 283$~MeV lattices to account for slight differences in the $\eta_c$ and $D$-meson masses when determined individually on these two volumes, following the approach described in Refs.~\cite{Wilson:2023hzu,Wilson:2023anv}.

\section{Finite volume spectra}

\begin{figure*}[t]
\includegraphics[width=0.99\textwidth]{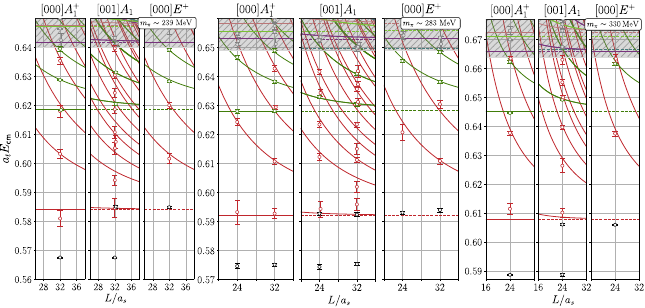}
\caption{ Finite-volume spectra in three irreps determined on the lattices described in Section~\ref{sec:lqcd}. Energy levels are colored according to their dominant operator overlap determined in the diagonalization of the matrix of correlation functions: $\eta_c \eta$-like (red), $D\bar{D}$-like (green), $c\bar{c}$-like (black). Grey points in the hatched high-energy region are not used in the subsequent determination of scattering amplitudes. Curves indicate the non-interacting spectrum of meson pairs on the relevant volumes with the same color-coding, with the addition of $\psi \omega$ (purple) and $D_s \bar{D}_s$ (light green) appearing above the energy cutoff. Dashed lines show the relevant meson-meson kinematic thresholds.
}
\label{fig:spec}
\end{figure*}

Figure~\ref{fig:spec} shows the discrete spectra of finite-volume energy levels on each of the lattices described in the previous section across three irreps. The $[000]\,A_1^+$ irrep at low energies will be dominated by $S$-wave scattering, while  $[000]\,E^+$ has its dominant contribution from $D$-wave scattering. $[001]\,A_1$ receives contributions from both $S$-wave and $D$-wave. As well as the finite-volume spectrum extracted from computed matrices of correlation functions, we also show the spectrum of \emph{non-interacting} meson pair energies in finite-volume, computed according to
\small
\begin{align*}
a_t E_\mathsf{lat}\! &=\! \Big[ (a_t m_1)^2 +  \big(\tfrac{2\pi}{\xi \, L/a_s}\big)^2 |\mathbf{n}_1|^2  \Big]^{1/2}\\
                     &\quad\quad+\! \Big[ (a_t m_2)^2 + \big(\tfrac{2\pi}{\xi\,  L/a_s}\big)^2 |\mathbf{n}_2|^2  \Big]^{1/2} \, , \\
a_t E_\mathsf{cm}\! &=\! \Big[ (a_t E_\mathsf{lat})^2 - \big(\tfrac{2\pi}{\xi \, L/a_s}\big)^2 |\mathbf{n}_2 + \mathbf{n}_1|^2\Big]^{1/2} \, ,
\end{align*}
\normalsize
where $\mathbf{n}_{1,2}$ are the discrete (integer triplet) momenta of the mesons which sum to the total momentum in the irrep characterization (see Refs.~\cite{Wilson:2023hzu,Wilson:2023anv} for more details appropriate to the current case).

The single low-lying energy level observed in $[000]A_1^+$, also seen as the lowest level in $[001]A_1$, is interpreted as being due to the stable $\chi_{c0}(1P)$, while the lowest energy level in $[000]E^+$, seen also in $[001]A_1$, is the stable $\chi_{c2}(1P)$. Above these we observe energy levels which are found to have dominant overlap onto either $\eta_c\eta$--like (red) or $D\bar{D}$--like (green) operators, all lying very close to non-interacting energies. 

The pattern of operator overlaps suggests limited channel-coupling between $\eta_c\eta$ and $D\bar{D}$ scattering channels, and the fact that the $D\bar{D}$-like states show no systematic departures from the $D\bar{D}$ non-interacting energies, leads us to anticipate only very weak, effectively decoupled, $D\bar{D}$ scattering at all three pion masses, and do not suggest the presence of any bound-states or resonances near to the $D\bar{D}$ threshold.

In the previous calculation with $m_\pi \sim 391$ MeV, reported on in Refs.~\cite{Wilson:2023hzu,Wilson:2023anv}, the spectrum was determined with a larger basis of operators allowing higher-lying levels to be reliably extracted. Within this extended spectrum, departures from non-interacting energies were observed higher in the spectrum, together with an `extra' level lying above the $D_s \bar{D}_s$ threshold, characteristic of a narrow resonance, as was found when the finite-volume spectrum was described by scattering amplitudes.

\section{Scattering amplitudes}

Much theoretical focus has been on the behavior of $D\bar{D}$ scattering in $S$-wave at kinematical threshold, but it is important to realize that this does not correspond to \emph{elastic} scattering, even in the approximation that charm-anticharm quark pairs are forbidden to annihilate. With the same quantum numbers, the two-hadron channel $\eta_c \eta$ is kinematically accessible. Some motivation for neglecting this channel comes from the longstanding empirically-motivated OZI-rule in which coupling of charmonia to \emph{open-charm} channels, like $D\bar{D}$, is expected to dominate over coupling to \emph{hidden-charm} channels, like $\eta_c \eta$, when both are kinematically open.

We wish to study this system in first-principles in QCD, so we will not make any unjustified assumptions about the relative importance of any particular open channel. Indeed, the L\"uscher approach, presented below, which relates scattering amplitudes to finite-volume spectra, requires us to consider all kinematically open scattering channels in the energy region of interest. In the current case, for all three pion masses considered, the $\eta_c \eta$ channel opens well below the $D\bar{D}$ channel, with the next channels able to contribute to $S$-wave scattering, $\psi \omega$ and $D_s \bar{D}_s$, opening at higher energy.
Our intent then is to determine coupled-channel $\etce - \DD$ \mbox{$S$-wave} amplitudes constrained by the finite-volume spectra presented in the previous section. From these amplitudes we will examine whether the channels are close to being decoupled, and whether there are strong interactions near the $D\bar{D}$ threshold which can be explained in terms of bound-states or resonances.

The relevant relationship between scattering amplitudes and finite-volume spectra~\cite{Luscher:1990ux, Hansen:2012tf,Briceno:2012yi}, sometimes referred to as the \emph{L\"uscher quantization condition}, can be written, 
\begin{align}
\det \left[\mathbf{1}+ i {\bm{\rho}(E)\cdot \bm{t}(E)\cdot \big(\bm{1}+ i \bm{\mathcal{M}}(E,L)\big)} \right]=0 \,,
\label{eq:luescher_det}
\end{align}
where $\bm{\rho}$ is a diagonal matrix of phase-space factors ${\rho_{ij}=(2k_i/\sqrt{s}) \, \delta_{ij}}$, $k_i$ is the $\cm$ momentum of the meson-meson pair in channel $i$, and $\bm{t}(E)$ is the scattering \mbox{$t$-matrix}. The kinematical matrix $\bm{\mathcal{M}}$ encapsulates the volume and irrep dependence in a known manner. Those energies where the determinant vanishes correspond to the spectrum, $E_\mathfrak{n}(L)$, in an $L\times L \times L$ periodic volume for a particular scattering amplitude $\bm{t}(E)$. 

Practical use of this equation to constrain coupled-channel scattering requires parameterization of the \mbox{$t$-matrix}, and this can be achieved in a manner that respects $s$-channel unitarity using the $K$-matrix, 
\begin{align*}
\bm{t}^{-1}=\bm{K}^{-1}+\bm{I} \, ,
\end{align*}
where $\mathrm{Im}\, \bm{I}=-\bm{\rho}$ above threshold. A real part for $\bm{I}$ can be generated from a dispersion relation if desired to improve somewhat the analytical properties of the $t$-matrix. For our purposes we will make use of the following simple  parameterizations of the $K$-matrix,
\begin{align}
K_{ij} &= (2k_i)^{\ell_i} \: \hat{K}_{ij} \: (2k_j)^{\ell_j}\, , \nonumber \\[1.2ex]
\hat{K}_{ij}&=\frac{g_i \, g_j}{m^2-s} +\gamma^{(0)}_{ij}+\gamma_{ij}^{(1)} \cdot \tfrac{s-4 m_D^2}{4 m_D^2} \, . 
\label{eq_Kmat_form}
\end{align}
The factors $(2k_i)^{\ell_i}$ are introduced to encourage the correct behavior at threshold for a partial-wave with angular momentum $\ell$. Inclusion of a pole-term in the parameterization allows for efficient description of a nearby bound-state or resonance, but in practice we will find that its only use in this calculation is to (optionally) describe the deeply-bound $\chi_{c0}(1P)$ and $\chi_{c2}(1P)$ states.

Ref.~\cite{Briceno:2017max} discusses the finite-volume approach in more detail, and Ref.~\cite{Woss:2020cmp} provides an efficient technique to solve Eq.~\ref{eq:luescher_det} in the coupled-channel case. As discussed in Refs.~\cite{Wilson:2023hzu,Wilson:2023anv}, the data correlation matrix characterizing the computed discrete energy levels can be rendered better conditioned by resetting a modest number of small singular values in computation of its inverse.

We will illustrate our analysis using the $m_\pi\sim283$~MeV lattice, on which we have the greatest degree of constraint, before discussing the two other pion masses, where the results are qualitatively similar.

\medskip
\subsection{$m_\pi\sim 283$~MeV}
\label{sec:lattice_levels_856}

As shown in Figure~\ref{fig:spec} (central panels), across two lattice volumes there are 42 energy-levels which can be used to constrain coupled-channel $\etce-\DD$ scattering in $S$-wave and $D$-wave below our choice of cutoff at $a_t E_\mathsf{cm} \sim 0.65$, lying somewhat below the $\psi \omega$ and $D_s \bar{D}_s$ thresholds. Eight of these levels (colored black in the figure) can be identified as being due to the stable $\chi_{c0}(1P)$ and $\chi_{c2}(1P)$ states, and in a first approximation these subthreshold bound-states can be assumed to have negligible impact upon the scattering system at higher energies. As such we will initially not attempt to describe them, and later we will relax this assumption, finding it to be very well justified.

We find that the remaining 34 energy levels can be described quite well by a \emph{constant} $K$-matrix for each of $S$-wave and $D$-wave, together with the dispersively improved Chew-Mandelstam phase space\footnote{Our implementation is described in Ref.~\cite{Wilson:2014cna}.} subtracted at the threshold energy for each channel,
\begin{widetext}
\footnotesize
  \begin{align*}
\renewcommand{\arraystretch}{1.54}
\begin{tabular}{rll}
$\gamma_{\eta_c\eta,\eta_c\eta}^{(S)} = $ & $(-0.09 \pm 0.20\pm 0.39)$ & \multirow{6}{*}{ $\begin{bmatrix*}[r]   1 &   \hphantom{-}0.39 &   \hphantom{-}0.15 & \quad -0.05 &  -0.01 &   0.07\\
&  1 &   0.01 &  -0.03 &  -0.04 &   0.03\\
&&  1 &   0.02 &  -0.00 &   0.11\\[2ex]
&&&  1 &   0.00 &  -0.09\\
&&&&  1 &  -0.01\\
&&&&&  1\end{bmatrix*}$ } \\ 
  $\gamma_{\eta_c\eta,\DD}^{(S)} = $  & $(0.00 \pm 0.12 \pm 0.19)$ & \\
  $\gamma_{\DD,\DD}^{(S)} = $         & $(0.66 \pm 0.27 \pm 0.35)$ & \\[2ex]
  $\gamma_{\etce,\etce}^{(D)} / a_t^4 = $     & $(1.0 \pm 0.5 \pm 1.3) \times 10^3$ & \\
  $\gamma_{\eta_c\eta,\DD}^{(D)}/ a_t^4 = $  & $(0.00 \pm 0.97 \pm 0.48) \times 10^3$ & \\
  $\gamma_{\DD,\DD}^{(D)}/ a_t^4 = $         & $(0.48 \pm 0.29 \pm 0.34) \times 10^3$
\end{tabular}
\end{align*}
\begin{align}
  \chi^2/ N_\mathrm{dof} = \tfrac{35.4}{34 - 6 - 3} =  1.41\,.
  \label{eq:fit_856_no_poles}
\end{align}
\end{widetext}

For each parameter, the first quoted uncertainty comes from the standard consideration of the curvature of the $\chi^2$ minimum, while the second reflects variation under adjustments of the scattering hadron masses and the anisotropy to their $\pm 1\sigma$ values. The correlations between parameters determined at the minimum are shown in the matrix on the right. The superscripts in circular brackets indicate the angular momentum $\ell$.
The reduced $\chi^2$ includes a subtraction of 3 degrees of freedom from 3 reset singular values in the computation of the inverse of the data correlation matrix\footnote{Retaining the full inverse data correlation matrix leads to extremely similar parameter values and a similar fit quality, $\chi^2/N_\mathrm{dof} = \tfrac{46.8}{34 - 4} =  1.56$. 
}.
Channel-coupling parameters are clearly being driven to zero value in this fit, and indeed forcing them to be exactly zero yields a slightly reduced value of $\chi^2/N_\mathrm{dof} = \tfrac{35.4}{34 - 4 - 3} =  1.31$. The remaining parameters, effectively describing \emph{decoupled} $\eta_c \eta - D\bar{D}$ scattering in $S$-wave and $D$-wave, are observed to be broadly compatible with zero, suggesting only weak scattering in each of these channels in the energy region constrained by our discrete energy levels. 

\begin{figure*}
\includegraphics[width=0.99\textwidth]{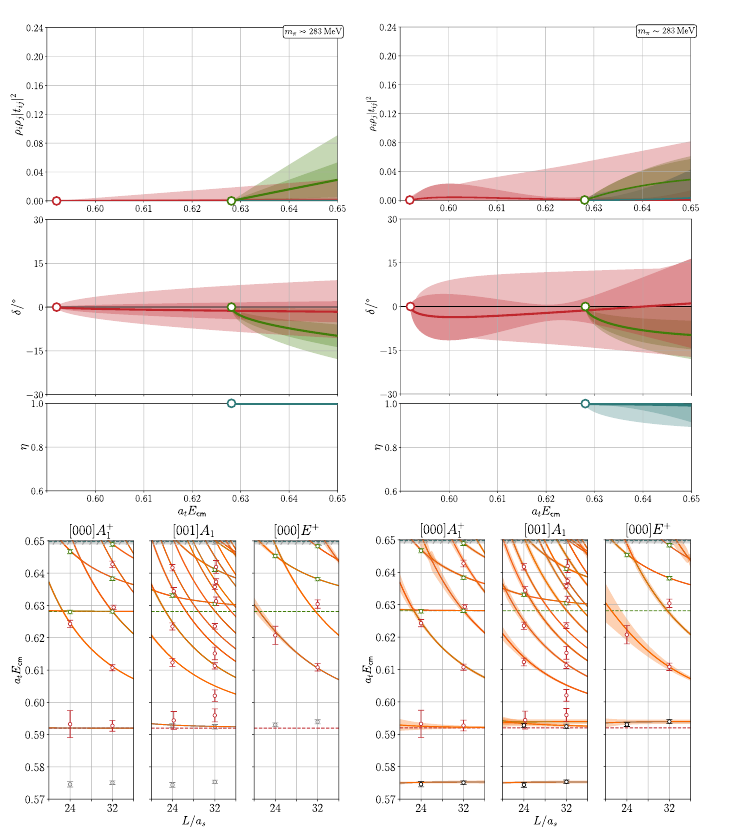}
\caption{Coupled $\eta_c \eta - D\bar{D}$ $S$-wave scattering amplitudes descriptions of finite-volume spectra. (Left) Amplitudes of Eq.~\ref{eq:fit_856_no_poles}. (Right) Amplitudes of Eq.~\ref{eq:fit_856_with_poles}.
Top panels illustrate the magnitude with a normalization where the unitarity limit is 1 for $t_{\eta_c \eta, \eta_c \eta}$ (red) and $t_{D\bar{D}, D\bar{D}}$ (green). 
Center panels show channel phase-shifts, $\delta_{\eta_c \eta}$ (red), $\delta_{D\bar{D}}$ (green) and inelasticity, $\eta$, defined so that the diagonal elements of the $2\times 2$ $t$-matrix are $t_{ii} = \tfrac{\eta e^{2 i \delta_i} - 1}{2i \rho_{i}}$ and the off-diagonal element is $t_{ij} = \tfrac{\sqrt{1-\eta^2} e^{i (\delta_i + \delta_j)}}{2 \sqrt{\rho_i \rho_j}}$.
Bottom panels show the finite-volume spectrum given by the amplitude (orange curves) compared with the spectrum constraining the amplitudes.
}
\label{fig:ref_amp_856}
\end{figure*}

This amplitude description is shown in Figure~\ref{fig:ref_amp_856} (left panel) where we observe explicitly the weak decoupled scattering in both $\eta_c \eta$ and $D\bar{D}$ channels. The bottom panel illustrates the good global description of the finite-volume spectrum provided by this amplitude, excluding the deeply bound $\chi_{c0}(1P)$ and $\chi_{c2}(1P)$ states that we did not attempt to describe.

The $S$-wave amplitude of Eq.~\ref{eq:fit_856_no_poles} has no nearby pole singularities on any of the four Riemann sheets of coupled-channel $\eta_c \eta - D\bar{D}$ scattering, as we would expect given the observed weakness of interaction.

\bigskip
Although we obtained an excellent description of the finite-volume spectrum above using a \emph{constant} $K$-matrix, we can also consider introducing more energy-dependence into the parameterization. In this case we choose to enforce $\eta_c \eta$, $D\bar{D}$ decoupling by setting off-diagonal elements of the $K$-matrix to zero, but we allow the $S$-wave $D\bar{D}$ \mbox{$K$-matrix} element to have a term in Eq.~\ref{eq_Kmat_form} with nonzero $\gamma^{(1)}$, i.e. allowing behavior linear in $s$, giving a fit description, 

{
\scriptsize
\begin{align*}
\renewcommand{\arraystretch}{1.54}
\begin{tabular}{rll}
$\gamma_{\eta_c\eta,\eta_c\eta}^{(0)(S)} = $ & $(-0.09 \pm 0.09 \pm 0.49)$ & \!\!\!\!\!\multirow{5}{*}{ $\begin{bmatrix*}[r]   1 &   0.14 &  \text{-}0.12 &   \quad 0.01 &   0.08\\
&  1 &  \text{-}0.95 &   0.22 &   0.42\\
&&  1 &  \text{-}0.23 &  \text{-}0.42\\[2ex]
&&&  1 &   0.01\\
&&&&  1\end{bmatrix*}$ } \\ 
$\gamma_{\DD,\DD}^{(0)(S)} = $ & $(-0.67 \pm 0.84 \pm 2.73)$ & \\
$\gamma_{\DD,\DD}^{(1)(S)} = $ & $(0.2 \pm 12.8 \pm 6.1)$ & \\[2ex]
$\gamma_{\eta_c\eta,\eta_c\eta}^{(0)(D)} / a_t^4 = $ & $(1.0 \pm 0.5 \pm 1.3) \!\times\! 10^3$ & \\
$\gamma_{\DD,\DD}^{(0)(D)} / a_t^4 = $ & $(0.47 \pm 0.32 \pm 0.32) \!\times\! 10^3$ & \\[1.3ex]
\end{tabular}
\end{align*}
\vspace{-0.4cm}
\begin{align}
  \chi^2/ N_\mathrm{dof} = \tfrac{35.4 }{34 - 5 - 3} =  1.36\,.
  \label{eq:fit_856_decoupled_linear}
\end{align}
}

We observe that the fit quality is essentially the same as the previous parameterization, with the coefficient of the term linear in $s$ being compatible with zero, and also significantly anticorrelated with the corresponding constant term. The resulting amplitudes show no systematic differences from  those plotted in Figure~\ref{fig:ref_amp_856} (left panel).

\bigskip
We can return to the question of whether the presence of the deeply-bound stable $\chi_{c0}(1P)$ and $\chi_{c2}(1P)$ states has any relevant impact for the scattering amplitudes above $\eta_c \eta$  threshold. Including the relevant finite-volume energy levels into our constraint, and adding a pole-term in each of the $S$-wave and the $D$-wave $K$-matrix parameterizations, we can find a description,
\begin{widetext}
\footnotesize
  \begin{align*}
\renewcommand{\arraystretch}{1.54}
\begin{tabular}{rll}
  $a_t\, m^{(S)} = $ & $(0.57528 \pm 0.00041 \pm 0.00002)$ & \multirow{9}{*}{ \renewcommand{\arraystretch}{1.54}
  $\begin{bmatrix*}[r]   1 &   \hphantom{-}0.70 &  -0.62 &   0.70 &  -0.70 &  \quad\quad 0.66 &   0.07 &  -0.05 &  -0.39\\
  &  1 &  -0.81 &   0.99 &  -1.00 &   0.94 &   0.10 &  -0.05 &  -0.50\\
  &&  1 &  -0.79 &   0.84 &  -0.78 &  -0.08 &   0.05 &   0.46\\
  &&&  1 &  -0.99 &   0.92 &   0.10 &  -0.06 &  -0.49\\
  &&&&  1 &  -0.93 &  -0.10 &   0.05 &   0.50\\[2ex]
  &&&&&  1 &   0.05 &  -0.04 &  -0.47\\
  &&&&&&  1 &  -0.15 &  -0.05\\
  &&&&&&&  1 &  -0.05\\
  &&&&&&&&  1\end{bmatrix*}$ } \\ 
  $a_t\, g_{\eta_c\eta}^{(S)} = $ & $(0.17 \pm 0.25 \pm 0.29)$ & \\
  $a_t\, g_{\DD}^{(S)} = $        & $(-0.28 \pm 0.10 \pm 0.10)$ & \\
  $\gamma_{\eta_c\eta,\eta_c\eta}^{(S)} = $ & $(0.4 \pm 1.5 \pm 0.0)$ & \\
  $\gamma_{\eta_c\eta,\DD}^{(S)} = $  & $(-0.7 \pm 2.5 \pm 0.0)$ & \\[2ex]
  $a_t\, m^{(D)} = $ & $(0.59392 \pm 0.00072 \pm 0.00000)$ & \\
  $g_{\eta_c\eta}^{(D)} / a_t = $ & $(4 \pm 9 \pm 55)^{(\dagger)}$ & \\
  $\gamma_{\etce,\etce}^{(D)} / a_t^4 = $ & $(1.3 \pm 0.5\pm 1.5) \times 10^3$ & \\
  $\gamma_{\DD,\DD}^{(D)} / a_t^4 = $ & $(0.37 \pm 0.26 \pm 0.28) \times 10^3 $ &
\end{tabular}
\end{align*}
\vspace{-0.4cm}
\begin{align}
  \chi^2/ N_\mathrm{dof} = \tfrac{32.1}{42 - 9 - 4} =  1.11\,.
  \label{eq:fit_856_with_poles}
\end{align}
\end{widetext}
%

Any parameter in Eq.~\ref{eq_Kmat_form} not specified above is fixed to zero in this description, and Chew-Mandelstam phase space is used, this time subtracted at the locations of the pole parameters $m^{(\ell)}$. 
Any effect above threshold from the presence of a \mbox{$D$-wave} pole below threshold is heavily suppressed by the centrifugal barrier such that it proves necessary to \emph{force} there to be a nonzero coupling to the $\eta_c \eta$ channel to ensure a solution to Eq.~\ref{eq:luescher_det} near the location of the $\chi_{c2}(1P)$ state. This is indicated by the $^{(\dagger)}$ above, where this parameter was limited in the $\chi^2$ minimization such that $|g_{\eta_c\eta}^{(D)}|/a_t \ge 4$. Using smaller magnitudes of this coupling cutoff, or forcing a nonzero coupling to $D\bar{D}$ instead, were also explored, with only tiny changes to the fit quality.

We observe that there are two well-determined \mbox{$K$-matrix} pole mass parameters in Eq.~\ref{eq:fit_856_with_poles}, whose numerical values closely correspond with the location of the low-lying previously excluded energy levels in Fig.~\ref{fig:spec}. As seen in Figure~\ref{fig:ref_amp_856} (right panel), the additional finite-volume energy levels below threshold are well described by this amplitude.

This description of the finite-volume spectrum can be seen in Figure~\ref{fig:ref_amp_856} (right panel) to again correspond to weak scattering above  threshold for both $\eta_c \eta$ and $D\bar{D}$, with no significant difference in behavior observed relative to the description ignoring the presence of the deeply-bound states.

The amplitude in Eq.~\ref{eq:fit_856_with_poles}, as expected, has an $S$-wave $t$-matrix pole at $a_t\sqrt{s}=0.57528(40)$ corresponding to the bound state $\ccz(1P)$. In addition there are further poles located  either on sheets not close to physical scattering, or distant in the complex energy plane, none of which have a significant influence on scattering above threshold.

The only relevant $t$-matrix pole in $D$-wave is found at $a_t\sqrt{s}=0.59392(76) - \frac{i}{2}\:0.00002(11)$ on the ${\mathrm{Im}\,  k_{\eta_c \eta} < 0}$, ${\mathrm{Im}\,  k_{D\bar{D}} > 0}$ Riemann sheet. That this state, corresponding to the $\chi_{c2}(1P)$, appears as a resonance with a tiny width is due to the forcing of a nonzero $\eta_c \eta$ coupling as described above -- this state can appear as a bound-state pole in amplitude variations that do not have this constraint.

\begin{figure}[h]
  \includegraphics[width=0.95\columnwidth]{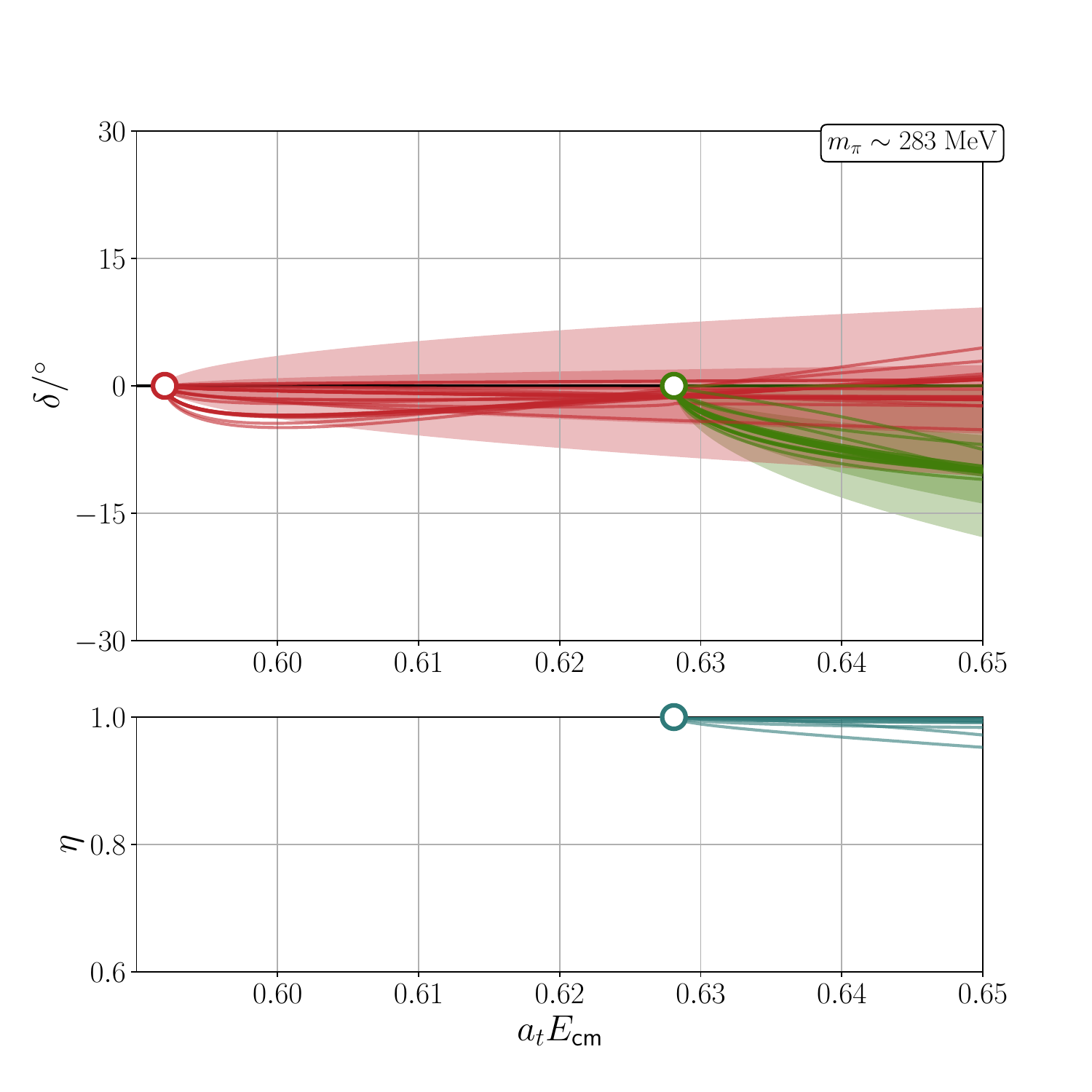}
  \caption{$S$-wave coupled-channel phase-shifts, $\delta_{\eta_c \eta}$, $\delta_{D\bar{D}}$, and inelasticity $\eta$, as defined in the caption of Figure~\ref{fig:ref_amp_856}, for a range of amplitude parameterizations capable of describing the finite-volume spectrum (Tables~\ref{tab:amp_variations_no_poles} and~\ref{tab:amp_variations}). 
For the amplitude given in Eq.~\ref{eq:fit_856_no_poles}, the uncertainty bands are shown (inner band is the statistical uncertainty, outer band includes also the impact of varying the masses and anisotropy as described in the text). For other amplitude parameterizations, only the central value curves are plotted.  
} 
  \label{fig:variations_dde_856}
\end{figure}

\medskip
In Figure~\ref{fig:variations_dde_856} we present $S$-wave amplitudes for a range of $K$-matrix parameterizations found to be capable of describing the finite-volume spectrum (presented in Appendix~\ref{app:amps}), where we observe that the finite-volume spectra have constrained quite tightly the weak magnitude of scattering, and the high degree of channel decoupling.


\subsection{$m_\pi \sim 239$~MeV and $m_\pi \sim 330$~MeV}

The lighter and heavier pion mass lattices have only a single volume each, and a smaller number of discrete finite-volume energies with which to constrain scattering amplitudes, as can be seen in Figure~\ref{fig:spec}. As in the ${m_\pi \sim 239}$ MeV case we limit our attention to energies lying below a cutoff slightly below the $\psi \omega$ threshold.


\begin{figure*}[!htb]
  \centering
  \begin{minipage}{0.99\columnwidth}
    \centering
    \includegraphics[width=\columnwidth]{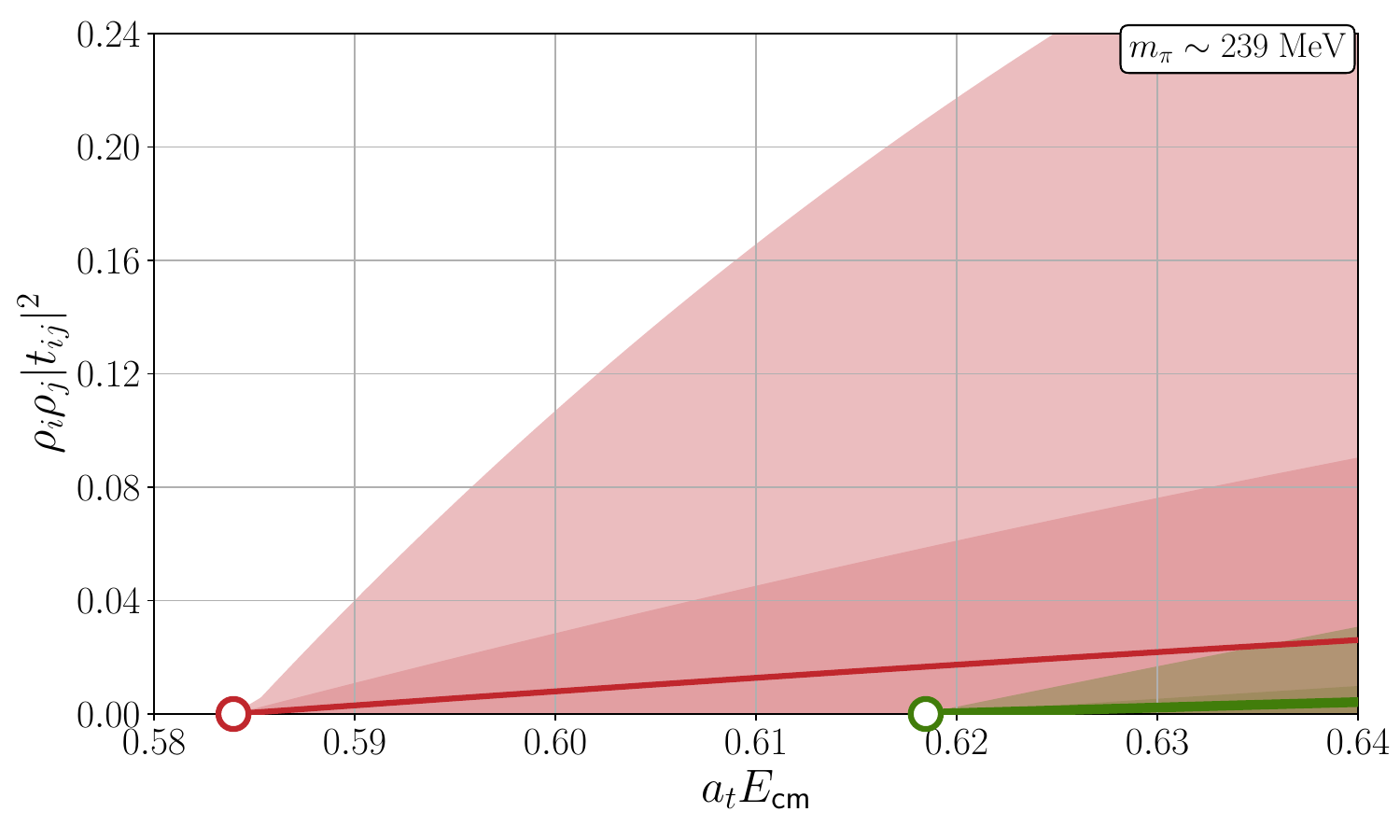}
  \end{minipage}\vspace{1em}
  \begin{minipage}{0.99\columnwidth}
    \centering
    \includegraphics[width=\columnwidth]{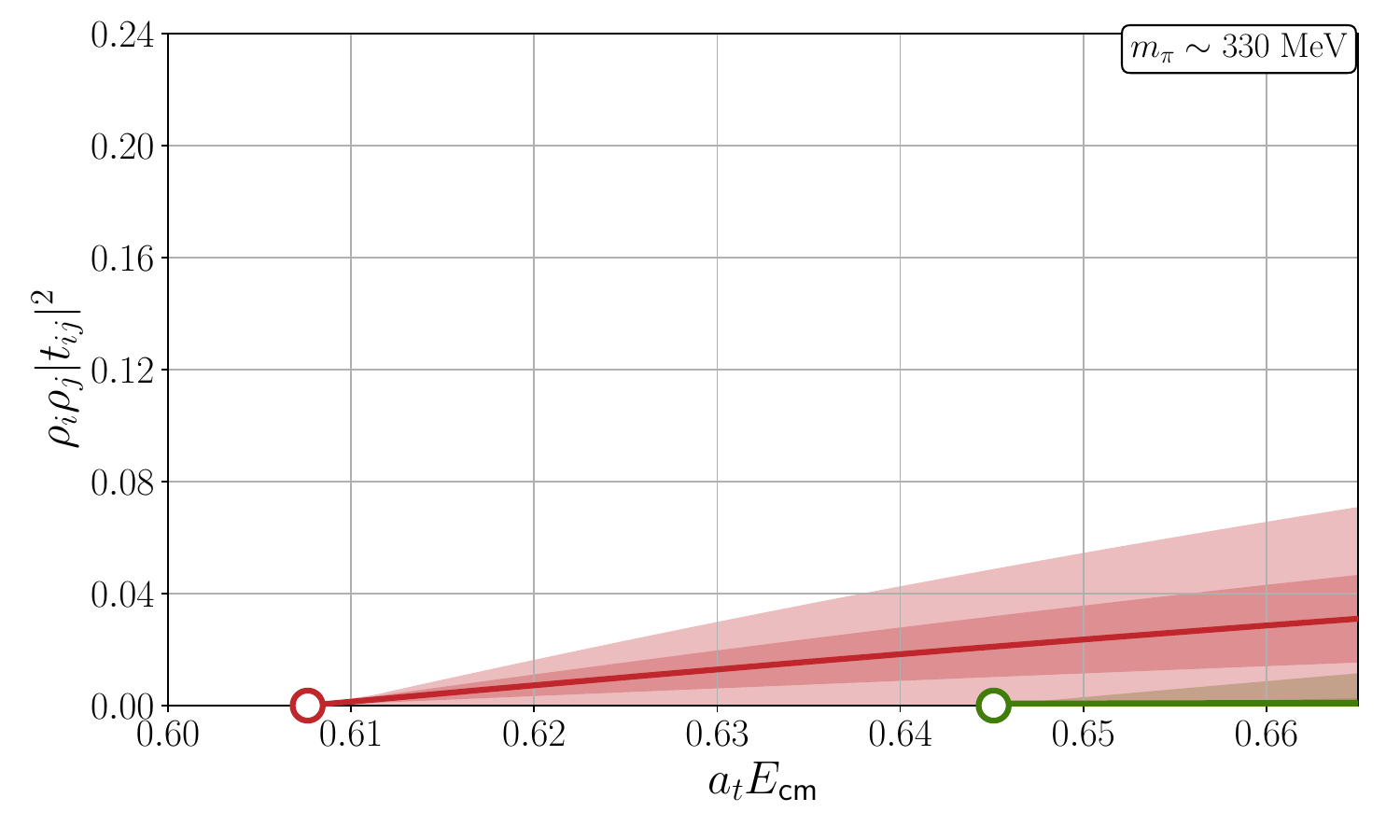}
  \end{minipage}
  \caption{$S$-wave scattering amplitudes for $m_\pi\sim 239$~MeV (left) and $330$~MeV (right). Same color coding as in top panels of Figure~\ref{fig:ref_amp_856}.}
  \label{fig:ref_amp_850_860}
\end{figure*}


In the $m_\pi \sim 239$~MeV case there are only five energy levels with dominant overlap onto $D\bar{D}$-like operators in the irreps dominated by $S$-wave scattering, which limits the degree to which we can consider amplitude parameterization variations.
As was the case for \mbox{$m_\pi \sim 283$ MeV} there are no obvious indications in the spectrum for strong interactions in either $\eta_c \eta$ or $D\bar{D}$, and a description in terms of a constant $K$-matrix with assumed $\eta_c \eta - D\bar{D}$ decoupling gives,
\begin{center}
\footnotesize
\renewcommand{\arraystretch}{1.5}
\begin{tabular}{rll}
$\gamma_{\eta_c\eta,\eta_c\eta}^{(S)} = $  & $(0.69 \pm 0.18\pm 0.74)$ & 
\!\!\!\!\!\!\!\!
\multirow{4}{*}{ 
\renewcommand{\arraystretch}{1.5}
$\begin{bmatrix*}[r]   1 &   \hphantom{\text{-}}0.42 &  \quad \text{-}1.00 &   0.02\\
&  1 &  \text{-}0.42 &   0.01\\[1ex]
&&  1 &  \text{-}0.02\\
&&&  1\end{bmatrix*}$ } \\ 
$\gamma_{\DD,\DD}^{(S)} = $ & $(0.07 \pm 0.19 \pm 0.00)$ & \\[1ex]
$\gamma_{\etce,\etce}^{(D)}/ a_t^4 = $ & $(\text{-}0.66 \pm 0.13 \pm 0.58) \!\times\! 10^3$ & \\
$\gamma_{\DD,\DD}^{(D)}/a_t^4 = $  & $(4.0 \pm 0.3 \pm 1.7) \!\times\! 10^3$ & \\[1.5ex]
\end{tabular}
\begin{align}
\chi^2/ N_\mathrm{dof} = \frac{11.4}{18-4-6} = 1.42\,,
  \label{eq:fit_860_no_poles}
\end{align}
\end{center}
and as shown in Figure~\ref{fig:ref_amp_850_860} (left) this corresponds to $S$-wave scattering that is compatible with zero for both $\eta_c \eta$ and $D\bar{D}$.

\bigskip

In the $m_\pi \sim 330$~MeV case, the use of a smaller volume leads to a relatively modest total number of energy levels, and only three energy levels with dominant overlap onto $D\bar{D}$--like operators in the irreps dominated by $S$-wave scattering. A description of the spectrum using a constant decoupled $K$-matrix, as above, yields,
\begin{center}
\footnotesize
\renewcommand{\arraystretch}{1.5}
\begin{tabular}{rll}
$\gamma_{\eta_c\eta,\eta_c\eta}^{(S)} = $ & $(0.58 \pm 0.16 \pm 0.30)$ & 
\!\!\!\!\!\!\!\!
\multirow{4}{*}{ 
\renewcommand{\arraystretch}{1.5}
$\begin{bmatrix*}[r]   1 &   0.32 &  \quad 0.44 &   0.20\\
&  1 &   0.21 &   0.52\\[1ex]
&&  1 &   0.13\\
&&&  1\end{bmatrix*}$ } \\ 
$\gamma_{\DD,\DD}^{(S)} = $ & $(-0.07\pm 0.28 \pm 0.20)$ & \\[1ex]
$\gamma_{\etce,\etce}^{(D)}/ a_t^4 = $ & $(0.34 \pm 0.13 \pm 0.60) \!\times\! 10^3$ & \\
$\gamma_{\DD,\DD}^{(D)}/a_t^4 = $ & $(0.27 \pm 0.22 \pm 0.11) \!\times\! 10^3$ & \\[1.5ex]
\end{tabular}
\begin{align}
\chi^2/ N_\mathrm{dof} = \frac{18.9}{15-4} = 1.72\,,
  \label{eq:fit_850_no_poles}
\end{align}
\end{center}
and as shown in Figure~\ref{fig:ref_amp_850_860} this again indicates scattering that is very weak in both $\eta_c \eta$ and $D\bar{D}$.

In summary, while we are not as heavily overconstrained in determination of scattering in the $m_\pi \sim 239$ MeV and 330 MeV cases as we were in the 283 MeV case, we still observe the same behavior, that of scattering which is compatible with being decoupled between $\eta_c \eta$ and $D\bar{D}$, and very weak in both. There are no indications for bound-states or resonances lying close to the $D\bar{D}$ threshold.

\section{Decoupled $D\bar{D}$ $S$-wave scattering as a function of $m_\pi$}

The amplitude descriptions of the previous section indicate no evidence for channel coupling between $D\bar{D}$ and $\eta_c \eta$ at any considered light quark mass, motivating a presentation of $D\bar{D}$ scattering near its kinematic threshold as an effectively \emph{elastic} process. It is common to plot \mbox{$S$-wave} elastic scattering straddling the threshold by means of the quantity $k \cot \delta_0$, which has the effective-range expansion, $a_0^{-1} + \tfrac{1}{2} r_0 k^2 + \ldots$, and which intersects $\pm \sqrt{- k^2}$ below threshold in the case of a bound-state (minus-sign choice) or a virtual-bound state (plus-sign choice). For the current case we have very weak scattering at threshold, characterized by small values of the scattering length, $a_0$, and as such a clearer presentation is given by plotting $k^{-1} \tan \delta_0$, which we may render dimensionless by scaling with the $D$-meson mass.

\begin{figure}[!ht]
  \includegraphics[width=0.9\columnwidth]{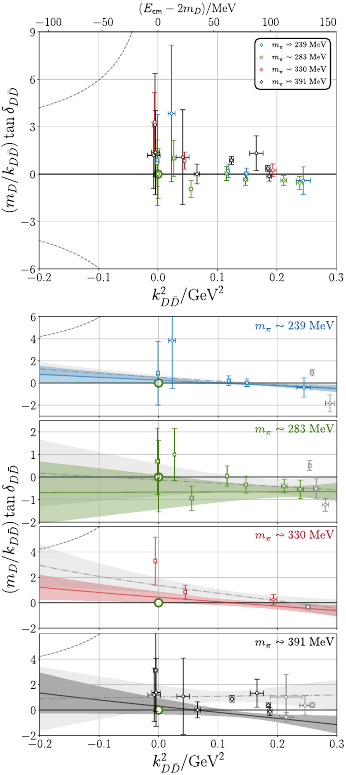}
  \caption{$\DD$ amplitudes plotted as $(m_D/k)\tan\delta$, with decoupled $S$-wave $\DD$ interactions. The large green circle shows the location of $\DD$ threshold. The energy level points selected have overlaps dominated by $\DD$ operators, and the $D$-wave has been fixed to zero in the quantization condition. The grey points are above the energy cutoff used in the fits for each lattice. The dashed grey curves on the left show $\pm m_D/|k|$ below threshold. We have removed 4 points from this plot for the $m_\pi\sim 391$~MeV case which have very large uncertainties. The grey bands and dot-dashed curves show amplitudes determined neglecting energy level correlations.}
  \label{fig:tandok}
\end{figure}

Figure~\ref{fig:tandok} shows the subset of energy levels in the $[000]A_1^+$ and $[001]A_1$ irreps which have dominant overlap with \mbox{$D\bar{D}$-like} operators. The relevant value of $\delta_0$ for each energy is computed assuming complete decoupling of the $\eta_c \eta$ channel, and zero amplitude for $D$-wave $D\bar{D}$ scattering at those energies. The upper panel indicates that there is no significant systematic dependence upon $m_\pi$ across the range from 239 MeV to 391 MeV. As well as the discrete points, the lower panels also show curves illustrating one particular parameterization considered in the previous section, fitted to the entire spectrum across all computed irreps, including $\eta_c \eta$-dominated levels\footnote{But excluding the $c\bar{c}(1P)$-identified levels. The particular amplitude choice is decoupled between $D\bar{D}$ and $\eta_c \eta$ with the $D\bar{D}$ $S$-wave described by a $K$-matrix element with both $\gamma^{(0)}$ and $\gamma^{(1)}$ terms, while the $D\bar{D}$ $D$-wave and the $\eta_c \eta$ $S$-wave and $D$-wave $K$-matrix elements are described by constants. The $m_\pi\sim 283$~MeV case (with spectrum correlations) is given in Eq.~\ref{eq:fit_856_decoupled_linear}.}.  These curves generally match the behavior of the discrete points, with the apparent discrepancy at intermediate energies in the $m_\pi \sim 391$ MeV case, and the systematically low fit in the $m_\pi \sim 330$ MeV case, being due to strong data correlations -- fits which neglect data correlation are also shown for comparison. We summarise these amplitudes in Appendix~\ref{app:decoupled_linear_amps}.

When plotted as $k^{-1} \tan \delta_0$ versus $k^2$, the presence of a bound-state would correspond to an intersection of the amplitude with the curve $-(-k^2)^{-1/2}$, while a virtual-bound-state would present through an intersection with $+(-k^2)^{-1/2}$. These curves are shown in Figure~\ref{fig:tandok} as dashed lines below threshold, and it is quite clear that the lattice QCD energy levels and amplitude descriptions following from them do not support any nearby bound-state or virtual-bound-state at any of the pion masses computed.


The strength of $D\bar{D}$ scattering at threshold can be characterized by an $S$-wave scattering length, $a_0^{D\bar{D}}$ defined as $\lim_{\sqrt{s} \to 2 m_D} t^{(\ell =0)}_{D\bar{D}, D\bar{D}}(s)$. In the case that $D\bar{D}$ was coupled to the open $\eta_c \eta$ channel the scattering length would take a complex value, but will be real if the channels are decoupled.
At all pion masses considered we find that for decoupled scattering amplitude descriptions we have $m_D\cdot |a_0^{D\bar{D}}| \lesssim 1$, with a slight preference for a small \emph{negative} value in the best-constrained case of \mbox{$m_\pi \sim 283$ MeV}. For comparison, the 4 MeV $D\bar{D}$ bound-state claimed in Ref.~\cite{Prelovsek:2020eiw} would correspond to a scattering length $m_D\cdot a_0^{D\bar{D}} \approx -22$.

\section{Interpretation}
\label{sec:interpretation}

The interpretation of the results of the calculation presented in this paper is quite simple: there is no evidence for strong scattering of $D\bar{D}$ in $S$-wave near threshold for pion masses between 239 MeV and 391 MeV. No pole singularities corresponding to bound-states or resonances are needed to describe the weak scattering observed. Refs.~\cite{Wilson:2023hzu,Wilson:2023anv} determined coupled-channel amplitudes at $m_\pi \sim 391$ MeV to higher energies, above the $D_s \bar{D}_s$ threshold, and found that a single narrow resonance was required, in line with the expectations of the $c\bar{c}$ quark model for a $2P$ state.

\medskip
A previous lattice QCD computation, Ref.~\cite{Prelovsek:2020eiw}, which used lattices on which $m_\pi \sim 280$ MeV, drew drastically different conclusions to those given above. 
This prior work chose to neglect the existence of the open $\eta_c \eta$ channel, computing correlation functions with $D\bar{D}$-like, $D_s \bar{D}_s$-like, and $c\bar{c}$-like operators\footnote{The lowest lying $D^* \bar{D}^*$-like and $J/\psi \,\omega$-like operators were also included.}. In light of our demonstration in this paper of complete decoupling with the $\eta_c \eta$ channel, this assumption is unlikely to have been a significant source of error.
Ref.~\cite{Prelovsek:2020eiw} concluded that their computed spectrum demanded the presence of a bound-state 4 MeV below the $D\bar{D}$ threshold, with this being largely driven by their lowest $D\bar{D}$-like levels appearing significantly below the lowest non-interacting energies in two irreps (e.g. $D\bar{D}$ threshold in the rest frame). Such an observation is not seen at any of the pion masses computed in this paper, where the lowest $D\bar{D}$-like levels lie extremely close to the non-interacting energies, and we find no place for such a bound state.

As well as the bound-state below $D\bar{D}$ threshold, and a `broad' resonance decaying to $D\bar{D}$, Ref.~\cite{Prelovsek:2020eiw} also proposed a nearly stable state sitting right at $D_s \bar{D}_s$ threshold, apparently almost completely decoupled from $D\bar{D}$. We have not explored amplitudes in this energy region directly in this paper, but it would be surprising if the $D_s \bar{D}_s$ system, containing no valence light quarks, depended strongly on the light quark mass, and no such near-stable state was found in the calculation at $m_\pi \sim 391$ MeV in Refs.~\cite{Wilson:2023hzu,Wilson:2023anv}. In the spectrum of higher-lying discrete energy levels computed in the current calculation for $m_\pi \sim 283$ MeV, but not used in amplitude determination, we do not see any obvious need for such a state.

A set of technical differences between the current calculation and that reported on in Ref.~\cite{Prelovsek:2020eiw} which could contribute to the difference in determined spectra are discussed in Appendices~\ref{app:prelovsek} and~\ref{app:corr_ratios}. A plausible candidate is the difference in lattice spacing which may lead to discretisation effects of differing magnitude. Within the calculation presented in this paper, done at (essentially) a single value of the lattice spacing, we see no evidence of discretization effects that impact our analysis, while in Ref.~\cite{Piemonte:2019cbi}, which used the same lattices as Ref.~\cite{Prelovsek:2020eiw}, significant discretization effects were highlighted.

\medskip
The current PDG~\cite{ParticleDataGroup:2024cfk} interpretation of the scalar resonance sector below 4 GeV, based upon the full set of contemporary experimental results, has a single `established' narrow resonance, $\chi_{c0}(3915)$, lying just below $D_s \bar{D}_s$ threshold, decaying to $D\bar{D}$ and $J/\psi\, \omega$ and contributing to a threshold peak in $D_s \bar{D}_s$ (observed in $B^+ \to K^+ \, D_s \bar{D}_s$ in Ref.~\cite{LHCb:2022aki}). Such a state could plausibly evolve into the single narrow resonance predicted at $m_\pi \sim 391$ MeV in Refs.~\cite{Wilson:2023hzu,Wilson:2023anv} as the light quark mass is increased\footnote{The resonance in Refs.~\cite{Wilson:2023hzu,Wilson:2023anv} lies above the $D_s \bar{D}_s$ threshold, but taking the amplitudes in Refs.~\cite{Wilson:2023hzu,Wilson:2023anv} and simply lowering slightly the $K$-matrix mass parameter such that the resonance lies just below the $D_s \bar{D}_s$ threshold would lead to a rapid turn on of $D_s \bar{D}_s$ amplitudes at threshold due to the large resonance coupling to that channel. Similar suggestions can be found in Refs.~\cite{Bayar:2022dqa,Guo:2022zbc}.}.

The only other scalar resonance candidate listed in the PDG, which is not considered to be `established', is the rather broad $\chi_{c0}(3860)$ claimed to decay into $D\bar{D}$ in $e^+e^- \to J/\psi \, D\bar{D}$~\cite{Belle:2017egg}. At the pion masses explored in this paper there is no sign of any amplitude behavior consistent with the presence of such a state.


\section{Summary}
 
We have explored the scattering of $D\bar{D}$ meson pairs in the region around the kinematic threshold, allowing for possible coupling to the open $\eta_c \eta$ channel, but finding that there is no evidence that the two channels couple at any significant level. Exploring a range of light quark masses corresponding to pion masses between 239 and \mbox{330 MeV} (supplemented by results previously obtained at 391 MeV in Refs.~\cite{Wilson:2023anv,Wilson:2023hzu}), we find no significant interaction strength is present. In particular the results suggest no evidence of a near-threshold bound, virtual or resonant $\ccz$ state.

Descriptions of the $D\bar{D}$ system at threshold  which demand a bound-state~\cite{Wang:2019evy,Wang:2020elp,Deineka:2021aeu,Dong:2021juy}, or a broader resonance (similar to the experimental candidate $\chi_{c0}(3860)$)~\cite{Guo:2012tv} are not supported by the results of this calculation.

An important caveat is that this work has been performed with only a single value of the gauge coupling, and therefore lattice spacings across the different light quark masses that are all close to a single value. While no obviously relevant indications of discretization effects have been observed, a systematic estimation of their scale is warranted in future. In addition, the operator basis used included none that resemble compact tetraquark constructions, however an earlier study using the same action and the same kind of optimised meson-meson operators suggest that these may not be required to obtain a reliable spectrum in this approach~\cite{Cheung:2016bym}.

{
This work indicates that isospin-0 $S$-wave scattering of $\DD$ close to threshold is \emph{much} simpler than the corresponding light-quark channels, $\pi\pi$ or $K\bar{K}$, with consistent behavior seen for a range of light-quark masses. Considered alongside the results presented in Refs.~\cite{Wilson:2023anv,Wilson:2023hzu} we have a demonstration, grounded in first-principles QCD, that the next scalar resonance above the $\chi_{c0}(1P)$ is likely to lie above 3900 MeV, with nothing of significance appearing in scattering amplitudes in between.
}

\acknowledgments

We thank our colleagues within the Hadron Spectrum Collaboration (\url{www.hadspec.org}), in particular Daniel Yeo for careful reading of the manuscript.
DJW acknowledges support from a Royal Society University Research Fellowship. DJW \& CET acknowledge support from the U.K. Science and Technology Facilities Council (STFC) [grant number ST/X000664/1].
JJD acknowledges support from the U.S. Department of Energy contract DE-SC0018416 at William \& Mary, and RGE from contract DE-AC05-06OR23177, under which Jefferson Science Associates, LLC, manages and operates Jefferson Lab. 
This work contributes to the goals of the U.S. Department of Energy ExoHad Topical Collaboration, Contract No. DE-SC0023598.

The software codes
{\tt Chroma}~\cite{Edwards:2004sx}, {\tt QUDA}~\cite{Clark:2009wm,Babich:2010mu}, {\tt QUDA-MG}~\cite{Clark:SC2016}, {\tt QPhiX}~\cite{ISC13Phi}, {\tt MG\_PROTO}~\cite{MGProtoDownload}, {\tt QOPQDP}~\cite{Osborn:2010mb,Babich:2010qb}, and {\tt Redstar}~\cite{Chen:2023zyy} were used. 
Some software codes used in this project were developed with support from the U.S.\ Department of Energy, Office of Science, Office of Advanced Scientific Computing Research and Office of Nuclear Physics, Scientific Discovery through Advanced Computing (SciDAC) program; also acknowledged is support from the Exascale Computing Project (17-SC-20-SC), a collaborative effort of the U.S.\ Department of Energy Office of Science and the National Nuclear Security Administration.

This work used the Cambridge Service for Data Driven Discovery (CSD3), part of which is operated by the University of Cambridge Research Computing Service (www.csd3.cam.ac.uk) on behalf of the STFC DiRAC HPC Facility (www.dirac.ac.uk). The DiRAC component of CSD3 was funded by BEIS capital funding via STFC capital grants ST/P002307/1 and ST/R002452/1 and STFC operations grant ST/R00689X/1. Other components were provided by Dell EMC and Intel using Tier-2 funding from the Engineering and Physical Sciences Research Council (capital grant EP/P020259/1). This work also used the earlier DiRAC Data Analytic system at the University of Cambridge. This equipment was funded by BIS National E-infrastructure capital grant (ST/K001590/1), STFC capital grants ST/H008861/1 and ST/H00887X/1, and STFC DiRAC Operations grant ST/K00333X/1. DiRAC is part of the National E-Infrastructure.
This work also used clusters at Jefferson Laboratory under the USQCD Initiative and the LQCD ARRA project.

Propagators and gauge configurations used in this project were generated using DiRAC facilities, at Jefferson Lab, and on the Wilkes GPU cluster at the University of Cambridge High Performance Computing Service, provided by Dell Inc., NVIDIA and Mellanox, and part funded by STFC with industrial sponsorship from Rolls Royce and Mitsubishi Heavy Industries. Also used was an award of computer time provided by the U.S.\ Department of Energy INCITE program and supported in part under an ALCC award, and resources at: the Oak Ridge Leadership Computing Facility, which is a DOE Office of Science User Facility supported under Contract DE-AC05-00OR22725; the National Energy Research Scientific Computing Center (NERSC), a U.S.\ Department of Energy Office of Science User Facility located at Lawrence Berkeley National Laboratory, operated under Contract No. DE-AC02-05CH11231; the Texas Advanced Computing Center (TACC) at The University of Texas at Austin; the Extreme Science and Engineering Discovery Environment (XSEDE), which is supported by National Science Foundation Grant No. ACI-1548562; and part of the Blue Waters sustained-petascale computing project, which is supported by the National Science Foundation (awards OCI-0725070 and ACI-1238993) and the state of Illinois. Blue Waters is a joint effort of the University of Illinois at Urbana-Champaign and its National Center for Supercomputing Applications.

\section*{Data access}

Reasonable requests for data, such as energy levels and correlations, can be directed to the authors and will be considered in accordance with the Hadron Spectrum Collaboration's policy on sharing data.

\bibliography{refs.bib}

\appendix

\section{Operator Basis}
\label{app:ops_lists}

In this section we present the basis of interpolating operators that were used to produce the matrices of correlation functions that form the input of our variational analyses. For $m_\pi \sim 283$ MeV see Table~\ref{tab:ops_856} and for $m_\pi \sim 239$ MeV and $m_\pi \sim 330$ MeV see Table~\ref{tab:ops_850_860}. Details of the operator basis can be found in Refs.~\cite{Wilson:2023hzu,Wilson:2023anv} and references therein.

\begin{table*}
\begin{tabular}{ccc}
\begin{tabular}{l|l}
\multicolumn{2}{c}{$[000]A_1^+$}\\[1ex]
$L/a_s=24$ & $L/a_s=32$ \\
\hline\\[-1.5ex]
$\bar{q}\Gamma q\times 7$  & $\bar{q}\Gamma q\times 7$ \\[1ex]
$\eta_c[000] \eta[000]$    & $\eta_c[000] \eta[000]$ \\
$\eta_c[001] \eta[001]$    & $\eta_c[001] \eta[001]$ \\
$\eta_c[011] \eta[011]$    & $\eta_c[011] \eta[011]$ \\
                           & $\eta_c[111] \eta[111]$ \\
                           & $\eta_c[002] \eta[002]$ \\[1ex]
$D[000] \bar{D}[000]$      & $D[000] \bar{D}[000]$ \\
$D[001] \bar{D}[001]$      & $D[001] \bar{D}[001]$ \\
                           & $D[011] \bar{D}[011]$ \\
                           & $D[111] \bar{D}[111]$ \\[1ex]
$D_s[000] \bar{D}_s[000]$  & $D_s[000] \bar{D}_s[000]$ \\[1ex]
$\psi[000] \omega[000]$    & $\psi[000] \omega[000]$ \\
\end{tabular}
\hspace{1cm}
&
\begin{tabular}{l|l}
\multicolumn{2}{c}{$[001]A_1$}\\[1ex]
$L/a_s=24$ & $L/a_s=32$ \\
\hline\\[-1.5ex]
$\bar{q}\Gamma q\times 10$ & $\bar{q}\Gamma q\times 11$ \\[1ex]
$\eta_c[001] \eta[000]$ & $\eta_c[001] \eta[000]$ \\
$\eta_c[000] \eta[001]$ & $\eta_c[000] \eta[001]$ \\
$\eta_c[011] \eta[001]$ & $\eta_c[011] \eta[001]$ \\
$\eta_c[002] \eta[001]$ & $\eta_c[002] \eta[001]$ \\
$\eta_c[001] \eta[011]$ & $\eta_c[001] \eta[011]$ \\
$\eta_c[111] \eta[011]$ & $\eta_c[111] \eta[011]$ \\
                        & $\eta_c[012] \eta[011]$ \\
                        & $\eta_c[011] \eta[111]$ \\
                        & $\eta_c[001] \eta[002]$ \\
                        & $\eta_c[112] \eta[111]$ \\[1ex]
$D[000] \bar{D}[001]$   & $D[000] \bar{D}[001]$ \\
$D[001] \bar{D}[011]$   & $D[001] \bar{D}[011]$ \\
                        & $D[001] \bar{D}[002]$ \\
                        & $D[011] \bar{D}[111]$ \\
\end{tabular}
\hspace{1cm}
&
\begin{tabular}{l|l}
\multicolumn{2}{c}{$[000]E^+$}\\[1ex]
$L/a_s=24$ & $L/a_s=32$ \\
\hline\\[-1.5ex]
$\bar{q}\Gamma q\times 10$ & $\bar{q}\Gamma q\times 8$ \\[1ex]
$\eta_c[001] \eta[001]$  & $\eta_c[001] \eta[001]$ \\
$\eta_c[011] \eta[011]$  & $\eta_c[011] \eta[011]$ \\
                         & $\eta_c[002] \eta[002]$ \\[1ex]
$D[001] \bar{D}[001]$    & $D[001] \bar{D}[001]$ \\
$D[011] \bar{D}[011]$    & $D[011] \bar{D}[011]$ \\[1ex]
$\psi[000] \omega[000]$  & $\psi[000] \omega[000]$ \\
\end{tabular}
\end{tabular}
\caption{Operators used in the $m_\pi \sim 283$~MeV variational analyses.}
\label{tab:ops_856}
\end{table*}

\begin{table*}
\begin{tabular}{l|l|l}
$[000]A_1^+$ & $[001]A_1$ & $[000]E^+$ \\[0.5ex]
\hline
$\bar{q}\Gamma q\times 7$ & $\bar{q}\Gamma q\times 14$ & $\bar{q}\Gamma q\times 11$ \\[1ex]
$\eta_c[000] \eta[000]$ & $\eta_c[001] \eta[000]$ & $\eta_c[001] \eta[001]$ \\
$\eta_c[001] \eta[001]$ & $\eta_c[000] \eta[001]$ & $\eta_c[011] \eta[011]$ \\
$\eta_c[011] \eta[011]$ & $\eta_c[011] \eta[001]$ & $\eta_c[002] \eta[002]$ \\
$\eta_c[111] \eta[111]$ & $\eta_c[002] \eta[001]$ &  \\
$\eta_c[002] \eta[002]$ & $\eta_c[001] \eta[011]$ &  \\
                        & $\eta_c[111] \eta[011]$ &  \\
                        & $\eta_c[012] \eta[011]$ &  \\
                        & $\eta_c[011] \eta[111]$ &  \\
                        & $\eta_c[001] \eta[002]$ &  \\
                        & $\eta_c[112] \eta[111]$ &  \\[1ex]
$D[000] \bar{D}[000]$        & $D[000] \bar{D}[001]$ & $D[001] \bar{D}[001]$        \\
$D[001] \bar{D}[001]$        & $D[001] \bar{D}[011]$ & $D[011] \bar{D}[011]$        \\
$D[011] \bar{D}[011]$        & $D[001] \bar{D}[002]$ & $D[002] \bar{D}[002]$        \\
$D[111] \bar{D}[111]$        & $D[011] \bar{D}[111]$ &         \\[1ex]
$D_s[000] \bar{D}_s[000]$  &                         & $D_s[001] \bar{D}_s[001]$ \\[1ex]
                           &                         & $D[011] \bar{D}^\ast[011]$ \\[1ex]
$\psi[000] \omega[000]$    &                         & $\psi[000] \omega[000]$  
\end{tabular}
\hspace{1.5cm}
\begin{tabular}{l|l|l}
$[000]A_1^+$ & $[001]A_1$ & $[000]E^+$ \\[0.5ex]
\hline
$\bar{q}\Gamma q\times 9$ & $\bar{q}\Gamma q\times 11$ & $\bar{q}\Gamma q\times 10$ \\[1ex]
$\eta_c[000] \eta[000]$ & $\eta_c[001] \eta[000]$ & $\eta_c[001] \eta[001]$ \\
$\eta_c[001] \eta[001]$ & $\eta_c[000] \eta[001]$ & $\eta_c[011] \eta[011]$ \\
$\eta_c[011] \eta[011]$ & $\eta_c[011] \eta[001]$ &  \\
                        & $\eta_c[002] \eta[001]$ &  \\
                        & $\eta_c[001] \eta[011]$ &  \\
                        & $\eta_c[111] \eta[011]$ &  \\
                        & $\eta_c[012] \eta[011]$ &  \\
                        & $\eta_c[011] \eta[111]$ &  \\[1ex]
$D[000] \bar{D}[000]$     & $D[000] \bar{D}[001]$     & $D[001] \bar{D}[001]$        \\
$D[001] \bar{D}[001]$     & $D[001] \bar{D}[011]$     & $D[011] \bar{D}[011]$        \\
                          & $D[011] \bar{D}[111]$      &                             \\[1ex]
$D_s[000] \bar{D}_s[000]$ & $D_s[000] \bar{D}_s[001]$ & $D_s[001] \bar{D}_s[001]$ \\[1ex]
                          &                           & $D[011] \bar{D}^\ast[011]$ \\[1ex]
                          &                           & $D^\ast[000] \bar{D}^\ast[000]$ \\[1ex]
$\psi[000] \omega[000]$   &                           & $\psi[000] \omega[000]$    
\end{tabular}
\caption{Operators used in the variational analyses. Left: $m_\pi \sim 239$~MeV, $L/a_s=32^3$. Right: $m_\pi \sim 330$~MeV, $L/a_s=24^3$.}
\label{tab:ops_850_860}
\end{table*}

\section{Amplitude Parameterization Variations for $m_\pi \sim 283$ MeV}\label{app:amps}

Tables~\ref{tab:amp_variations_no_poles} and \ref{tab:amp_variations} show the various parameterizations used to describe the finite-volume spectra excluding (Table~\ref{tab:amp_variations_no_poles}) and including (Table~\ref{tab:amp_variations}) the $c\bar{c}(1P)$ deeply-bound energy levels. The resulting amplitudes are plotted in Figure~\ref{fig:variations_dde_856}.

\begin{table*}
\renewcommand{\arraystretch}{1.2}
{\scriptsize
\begin{tabular}{cccccc|ccc|cc|r}
\multicolumn{6}{c|}{$S$-wave parameters} &
\multicolumn{3}{c|}{$D$-wave parameters} & 
\multicolumn{2}{c|}{Phase space} &
\multicolumn{1}{c}{
\multirow{2}{*}{$\dfrac{\chi^2}{N_\mathrm{dof}}$}
 } \\
$\gamma_{1,1}^{(S)(0)}$ & $\gamma_{1,1}^{(S)(1)}$ & $\gamma_{1,2}^{(S)(0)}$ & $\gamma_{1,2}^{(S)(1)}$ & $\gamma_{2,2}^{(S)(0)}$ & $\gamma_{2,2}^{(S)(1)}$ &
$\gamma_{1,1}^{(D)(0)}$ & $\gamma_{1,2}^{(D)(0)}$ & $\gamma_{2,2}^{(D)(0)}$ & 
$S$ & $D$ \\
\hline
\\
\multicolumn{12}{l}{Reference parameterisation Eq.~\ref{eq:fit_856_no_poles}:}\\
$\tick$ & $-$ & $\tick$ & $-$ & $\tick$     & $-$ &
$\tick$ & $\tick$ & $\tick$ &
CM & CM & $\tfrac{35.4}{34 - 6 - 3} =  1.41$\\
\hline
\\
\multicolumn{12}{l}{Parameterisation variations using Eq.~\ref{eq_Kmat_form} with no poles:}\\
$\tick$ & $-$ & $\tick$ & $-$ & $\tick$     & $-$ &
$\tick$ & $-$ & $\tick$     & 
CM & CM & $\tfrac{35.4}{34 - 5 - 3} =  1.36$\\
$\tick$ & $-$ & $\tick$ & $-$ & $-$     & $-$ &
$\tick$ & $\tick$ & $-$     & 
CM & CM & $\tfrac{35.4}{34 - 4 - 3} =  1.31$\\
$\tick$ & $-$ & $-$     & $-$ & $\tick$     & $-$ &
$\tick$ & $-$ & $\tick$     & 
CM & CM & $\tfrac{35.4}{34 - 4 - 3} =  1.31$\\
$\tick$ & $-$ & $-$     & $\tick$ & $\tick$     & $-$ &
$\tick$ & $-$ & $\tick$     & 
CM & CM & $\tfrac{35.4}{34 - 5 - 3} =  1.36$\\
$\tick$ & $-$ & $-$     & $-$ & $\tick$     & $\tick$ &
$\tick$ & $-$ & $\tick$     & 
CM & CM & $\tfrac{35.4}{34 - 5 - 3} =  1.36$\\
$\tick$ & $-$ & $-$     & $-$ & $-$     & $\tick$ &
$\tick$ & $-$ & $\tick$     & 
CM & CM & $\tfrac{35.5}{34 - 4 - 3} =  1.31$\\
$-$ & $\tick$ & $-$     & $-$ & $\tick$     & $-$ &
$\tick$ & $-$ & $\tick$     & 
CM & CM & $\tfrac{35.4}{34 - 4 - 3} =  1.31$\\
$\tick$ & $-$ & $\tick$ & $-$ & $\tick$     & $-$ &
$\tick$ & $\tick$ & $\tick$     & 
$-\mathrm{i}\rho$ & $-\mathrm{i}\rho$ & $\tfrac{35.4}{34 - 6 - 3} =  1.42$\\
$\tick$ & $-$ & $-$     & $-$ & $\tick$     & $-$ &
$\tick$ & $-$ & $\tick$     & 
$-\mathrm{i}\rho$ & $-\mathrm{i}\rho$ & $\tfrac{35.4}{34 - 4 - 3} =  1.31$\\
\hline
\end{tabular}
}
\caption{Parameterization variations. Channel labels: 1 $\to \etce$, $2\to\DD$, CM: Chew-Mandelstam, $-\mathrm{i}\rho$: simple phase space.}
\label{tab:amp_variations_no_poles}
\end{table*}

\begin{table*}
\renewcommand{\arraystretch}{1.2}
{\scriptsize
\begin{tabular}{cccccc|cccccc|cc|r}
\multicolumn{6}{c|}{$S$-wave parameters} &
\multicolumn{6}{c|}{$D$-wave parameters} & 
\multicolumn{2}{c|}{Phase space} &
\multicolumn{1}{c}{
\multirow{2}{*}{$\dfrac{\chi^2}{N_\mathrm{dof}}$}
 } \\
$m_S$ & $g_{1}^{(S)}$ & $g_{2}^{(S)}$ & $\gamma_{1,1}^{(S)}$ & $\gamma_{1,2}^{(S)}$ & $\gamma_{2,2}^{(S)}$ & 
$m_D$ & $g_{1}^{(D)}$ & $g_{2}^{(D)}$ & $\gamma_{1,1}^{(D)}$ & $\gamma_{1,2}^{(D)}$ & $\gamma_{2,2}^{(D)}$ &
$S$ & $D$ \\
\hline
\\
\multicolumn{15}{l}{Parameterisation from Eq.~\ref{eq:fit_856_with_poles}:}\\
$\tick$ & $\tick$ & $\tick$     & $\tick$ & $\tick$     & $-$ &
$\tick$ & $\tick$ & $-$     & $\tick$ & $-$ & $\tick$ &
CM & CM & $\tfrac{32.1}{42 - 9 - 4} =  1.11$\\
\hline
\\
\multicolumn{15}{l}{Parameterisation variations using Eq.~\ref{eq_Kmat_form} (including poles):}\\
$\tick$ & $\tick$ & $\tick$ & $\tick$ & $\tick$     & $\tick$ &
$\tick$ & $\tick$ & $\tick$     & $\tick$ & $\tick$ & $\tick$ &
CM & CM & $\tfrac{32.1}{42 - 12 - 4} =  1.23$\\
$\tick$ & $\tick$ & $\tick$ & $\tick$ & $\tick$     & $\tick$ &
$\tick$ & $\tick$ & $\tick$     & $\tick$ & $-$ & $\tick$ &
CM & CM & $\tfrac{32.1}{42 - 11 - 4} =  1.19$\\
$\tick$ & $\tick$ & $-$ & $\tick$ & $\tick$     & $\tick$ &
$\tick$ & $\tick$ & $\tick$     & $\tick$ & $-$ & $\tick$ &
CM & CM & $\tfrac{32.1}{42 - 10 - 4} =  1.14$\\
$\tick$ & $\tick$ & $\tick$ & $\tick$ & $\tick$     & $\tick$ &
$\tick$ & $\tick$ & $-$     & $\tick$ & $-$ & $\tick$ &
CM & CM & $\tfrac{32.1}{42 - 10 - 4} =  1.14$\\
$\tick$ & $\tick$ & $\tick$ & $\tick$ & $\tick$     & $-$ &
$\tick$ & $\tick$     & $-$     & $-$     & $\tick$ & $\tick$ &
CM & CM & $\tfrac{38.5}{42 - 9 - 4} =  1.33$\\
$\tick$ & $\tick$ & $\tick$ & $\tick$ & $-$     & $-$ &
$\tick$ & $\tick$     & $-$     & $-$     & $\tick$ & $\tick$ &
CM & CM & $\tfrac{39.6}{42 - 8 - 4} =  1.32$\\
$\tick$ & $\tick$ & $\tick$ & $-$ & $-$     & $-$ &
$\tick$ & $\tick$     & $-$     & $-$     & $\tick$ & $\tick$ &
CM & CM & $\tfrac{39.7}{42 - 7 - 4} =  1.28$\\
$\tick$ & $\tick$ & $-$ & $\tick$ & $\tick$     & $\tick$ &
$\tick$ & $\tick$     & $-$     & $-$     & $\tick$ & $\tick$ &
CM & CM & $\tfrac{39.9}{42 - 9 - 4} =  1.38$\\
$\tick$ & $\tick$     & $-$     & $\tick$     & $-$ & $\tick$ &
$\tick$ & $\tick$     & $-$     & $\tick$     & $-$ & $\tick$ &
CM & CM & $\tfrac{ 32.1}{42 - 8 - 4} =  1.07$\\
\\
$\tick$ & $\tick$     & $-$     & $\tick$     & $-$ & $\tick$ &
$\tick$ & $\tick$     & $-$     & $\tick$     & $-$ & $\tick$ &
$-\mathrm{i}\rho$ & $-\mathrm{i}\rho$ & $\tfrac{32.1}{42 - 8 - 4} =  1.07$\\
$\tick$ & $\tick$ & $\tick$ & $\tick$ & $\tick$     & $\tick$ &
$\tick$ & $\tick$ & $\tick$     & $\tick$ & $\tick$ & $\tick$ &
$-\mathrm{i}\rho$ & $-\mathrm{i}\rho$  & $\tfrac{32.1}{42 - 12 - 4} =  1.23$\\
$\tick$ & $\tick$ & $\tick$ & $\tick$ & $\tick$     & $-$ &
$\tick$ & $\tick$ & $-$     & $\tick$ & $-$ & $\tick$ &
$-\mathrm{i}\rho$ & $-\mathrm{i}\rho$  & $\tfrac{ 32.2}{42 - 9 - 4} =  1.11$\\
$\tick$ & $\tick$ & $\tick$ & $-$ & $-$     & $-$ &
$\tick$ & $\tick$     & $-$     & $-$     & $\tick$ & $\tick$ &
$-\mathrm{i}\rho$ & $-\mathrm{i}\rho$ & $\tfrac{39.7}{42 - 7 - 4} =  1.28$\\
\hline
\end{tabular}
}
\caption{Parameterization variations. Channel labels: 1 $\to \etce$, $2\to\DD$, CM: Chew-Mandelstam, $-\mathrm{i}\rho$: simple phase space.}
\label{tab:amp_variations}
\end{table*}


\section{$\DD$ amplitudes from decoupled linear $K$-matrix parameterisations, appearing in Fig.~\ref{fig:tandok}}
\label{app:decoupled_linear_amps}

In Table~\ref{tab:linear_DD_amps}, we summarise the $S$-wave amplitudes presented in Fig.~\ref{fig:tandok}. For each light quark mass, we consider both a correlated $\chi^2$ fit and, for visual comparison, one where the energy correlations are discarded. The amplitude form is the same as that used in Eq.~\ref{eq:fit_856_decoupled_linear}, as defined in Eq.~\ref{eq_Kmat_form}, with a Chew-Mandelstam phase-space that is subtracted at $D\bar{D}$ threshold.  With this fit-form the fit parameter $\gamma^{(0)}_{\DD}$ is equal to $m_D\, a^{\DD}_{\ell=0}$, where $a^{\DD}_{\ell=0}$ is the $S$-wave $\DD$ scattering length. 

We include fits with an uncorrelated $\chi^2$ to illustrate that the apparent systematic difference between the points and curves in Fig.~\ref{fig:tandok} are due to data correlation -- when these correlations are ignored the fit curves are in good visual agreement with the points. Both correlated and uncorrelated fits indicate very weak interactions at threshold.

\begin{table}
\renewcommand{\arraystretch}{1.4}
\begin{tabular}{lrrrr}
& $\quad$ 239 MeV & $\quad$ 283 MeV & $\quad$ 330 MeV & $\quad$ 391 MeV \\[0.2ex]
\hline
\multicolumn{4}{l}{\emph{correlated}}\\
$\gamma^{(0)}_{\DD}$             & $0.18 \pm 0.15$ & $-0.67 \pm 0.84$  & $0.44\pm 0.47$       & $0.29 \pm 0.63$ \\
$\gamma^{(1)}_{\DD}$             & $-9.1 \pm 1.1$   & $0.2 \pm 12.8$    & $-12.5 \pm 9.0$     & $-17 \pm 14$ \\
$\chi^2/N_\mathrm{dof}$          & 1.75             & 1.36              & 1.70                & 1.37\\
%
%
\hline
\multicolumn{4}{l}{\emph{uncorrelated}}\\
$\gamma^{(0)}_{\DD}$             & $0.51 \pm 0.20$  & $-0.13 \pm 0.96$ & $1.38\pm 0.60$    & $1.02 \pm 0.94$\\
$\gamma^{(1)}_{\DD}$             & $-11.9 \pm 6.0$ & $-5 \pm 17$      & $-22 \pm 14$      & $3 \pm 26$\\
$\chi^2/N_\mathrm{dof}$          & 0.95             & 1.15             & 1.16              & 0.84\\[0.2ex]
%
%
\hline
\end{tabular}
\caption{$S$-wave $\DD$ amplitudes determined using a linear decoupled two-channel $K$-matrix that are shown in Fig.~\ref{fig:tandok}. In the definitions used here $m_Da^{\DD}_{\ell=0} = \gamma^{(0)}_{\DD}$.}
\label{tab:linear_DD_amps}
\end{table}

\section{Additional details and comparison of results with Prelovsek \emph{et al.}, Ref.~\cite{Prelovsek:2020eiw}}
\label{app:prelovsek}

Ref.~\cite{Prelovsek:2020eiw} concludes a completely different spectrum of bound-states and resonances, and strength of $D\bar{D}$ scattering to the current calculation, even when the quark mass is comparable at $m_\pi \sim 280$ MeV. These systematic differences with respect to Ref.~\cite{Prelovsek:2020eiw} do not have an obvious explanation.
The relatively large energy shifts with respect to finite-volume non-interacting energies found in Ref.~\cite{Prelovsek:2020eiw} are what drive their amplitudes to contain the extra singularities, and as such we should attempt to identify what is causing the finite-volume spectrum to differ so significantly.

\medskip
One possibility is that the extracted finite-volume energy levels in one or both studies are systematically biased, or have an underestimated systematic error. Their determination in principle depends upon the details of operator construction, the content of the operator basis, the procedure used to diagonalize the correlator matrix, and the timeslice fitting of the correlator matrix eigenvalues. We will highlight some differences in approach between the calculation presented in this paper, and that in Ref.~\cite{Prelovsek:2020eiw}, although we lack sufficient information about the details of analysis used in Ref.~\cite{Prelovsek:2020eiw} to make a full quantitative comparison.

In the current study, we make use of \emph{optimized} single-hadron operators in the construction of hadron-hadron operators. They are optimized in the sense of having been determined by diagonalizing a matrix of correlation functions built using a basis of fermion bilinear operators, independently for each meson three-momentum~\cite{Dudek:2012gj}. Using these optimized operators in the pair-products of hadron-hadron operators, we might expect more rapid relaxation to the finite-volume energies given the suppression of single-hadron excitations. 
In Refs.~\cite{Prelovsek:2025vbr,Stump:2025owq}, differences in extracted finite-volume energy level values are seen when adding operators formed from a diquark-antidiquark construction to hadron-hadron constructions, when compared to using hadron-hadron constructions alone. Like Ref.~\cite{Prelovsek:2020eiw}, these studies use hadron-hadron operator constructions that use non-optimized single-hadron operators in the product.

\begin{figure*}
  \includegraphics[width=\textwidth]{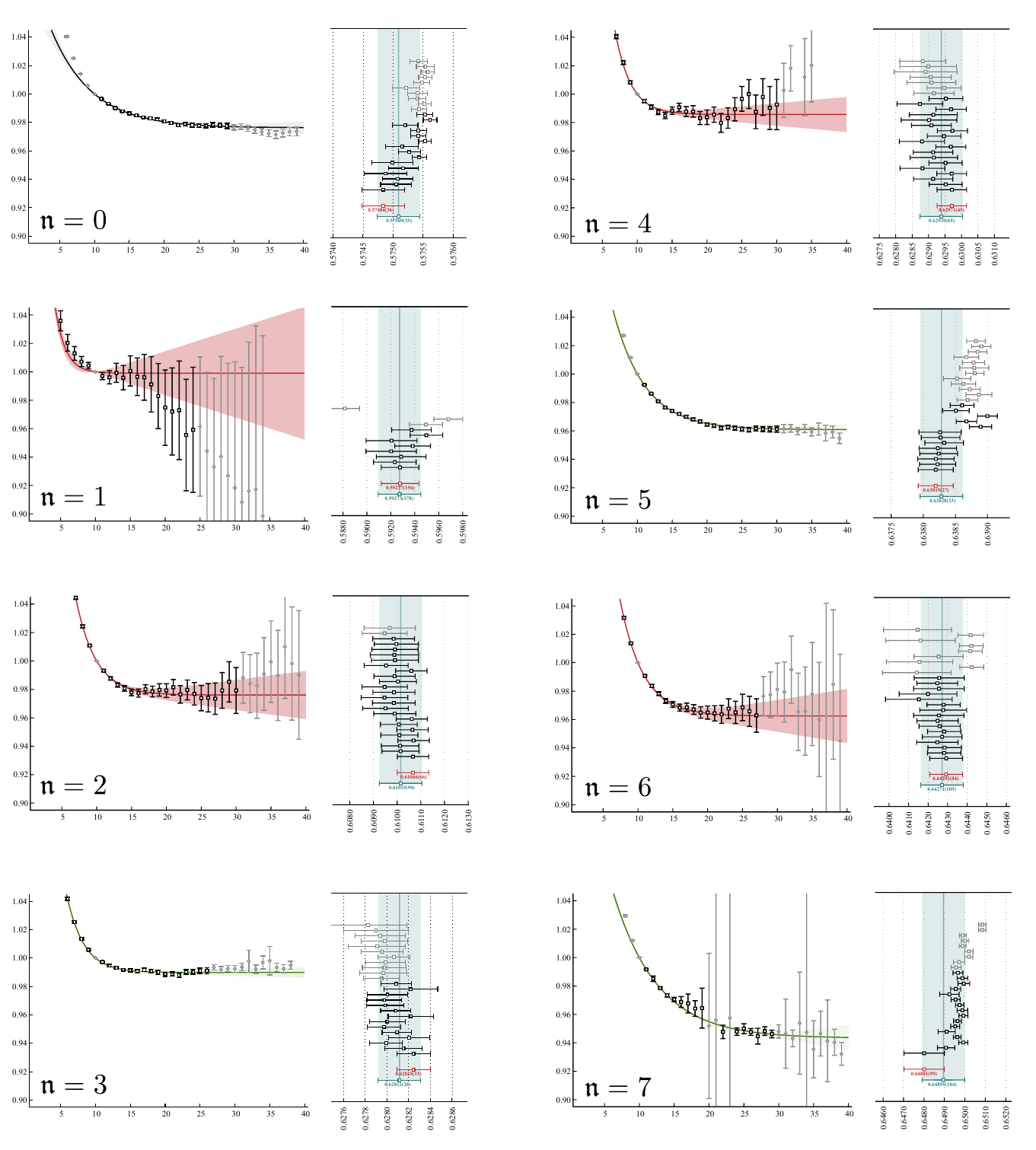}
  \caption{Timeslice fits to \emph{principal correlators}, $\lambda_\mathfrak{n}(t)$, obtained from solving generalized eigenvalue problem, $\mathbf{C}(t) v_\mathfrak{n} = \lambda_\mathfrak{n}(t) \mathbf{C}(t_0) v_\mathfrak{n}$, with $t_0/a_t = 10$. Fit with $\lambda_\mathfrak{n}(t) = (1-A_\mathfrak{n}) e^{-E_\mathfrak{n} (t-t_0)} + A_\mathfrak{n} e^{-E_\mathfrak{n}' (t-t_0)}$. $m_\pi \sim 283$ MeV, $32^3$ lattice, $[000]\, A_1^+$ irrep. Each panel shows, on left, $e^{E_\mathfrak{n} (t-t_0)} \,\lambda_\mathfrak{n}(t)$, with the best single time-window fit curve superimposed, and on right the variation of extracted $a_t E_\mathfrak{n}$ over different fit windows, ordered from bottom up by decreasing value of AIC. Energy values with AIC probability $ < 1\%$ are greyed out. The AIC model average value of $a_t E_\mathfrak{n}$ is shown by the blue point and band.
  }
  \label{fig:princorrs}
\end{figure*}

Figure~\ref{fig:princorrs} shows an example of the extraction of a finite-volume spectrum from the generalized eigenvalues of the matrix of correlation functions. The example shown is the $[000]\, A_1^+$ irrep on the $32^3$ lattice at $m_\pi \sim 283$ MeV. For each eigenvalue at $t_0 = 10 \, a_t$ we see behavior consistent with dominance of a single state at modest times. Each time-dependence is fitted as a sum of two exponentials (and occasionally as a single exponential) for a range of time-windows, and each fit has a quality associated with it as estimated by the Akaike Information Criterion (AIC). Our final determination for each $a_t E_\mathfrak{n}$ value comes from the `model average' over these determinations (see Ref.~\cite{Jay:2020jkz}). As can be seen from the blue points (AIC average) compared to the red points (single best time-window fit), this average typically provides a more conservative estimate.
We note that the near-threshold $D\bar{D}$-operator dominated state (bottom left panel) has a principal correlator which very clearly relaxes to its long-time behavior rapidly and shows very little variation over time-window choice.

\begin{figure}
  \includegraphics[width=0.7\columnwidth]{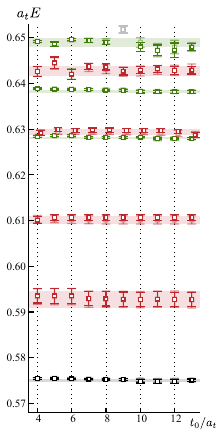}
  \caption{In same case as Figure~\ref{fig:princorrs}, the variation of extracted $E_\mathfrak{n}$ spectrum with changing choice of $t_0$. Solid points show the single best fit-window value, and the ghosted values show the AIC model average. Horizontal bands illustrate the $t_0 = 10\, a_t$ choice.
  }
  \label{fig:t0}
\end{figure}

Figure~\ref{fig:t0} shows that the choice of $t_0$ value used in the solution of the generalized eigenvalue problem in this case has no significant impact upon the extracted energy spectrum.


\medskip
Potentially important differences between the two calculations come from the details of the lattice setup, the lattice spacing, and the implementation of charm quarks.
Ref.~\cite{Prelovsek:2020eiw} uses isotropic lattices with $a \approx 0.086 \, \mathrm{fm}$, and while this is finer than our spatial spacing, it is much coarser than our temporal spacing. Using clover improved Wilson fermions to describe charm quarks, they report evidence for discretization effects in the form of departures from continuum relativistic dispersion relations for $D$-mesons.
They choose to apply an \emph{ad-hoc} correction to deal with this in their finite-volume energy spectrum: for each determined energy level, after computing the energy gap from the nearest non-interacting $D\bar{D}$ energy (using $D$-meson energies as computed on the lattice) they then produce an absolute energy by reinstating the non-interacting energy after moving the $D$-meson energies to lie on the continuum dispersion relation. No additional systematic error is associated with this procedure, despite the adjustment in energy being typically larger than the statistical error on the energy level. This procedure impacts higher-lying energies, but does not significantly shift the levels at threshold in the rest-frame, indicating that it alone cannot be the source of the differences with respect to the energy levels in the current paper.

In comparison, in the calculation reported on in this paper, we observe no systematic departures from the continuum relativistic dispersion form, \emph{but} we have the additional anisotropy parameter, $\xi$, to vary, and as presented in Sec.~\ref{sec:lqcd} this parameter varies slightly for different hadron species. We treat this variation conservatively by propagating the maximal deviations into our final amplitudes (see Table~\ref{tab:xi}).

\begin{figure}
  \includegraphics[width=0.9 \columnwidth]{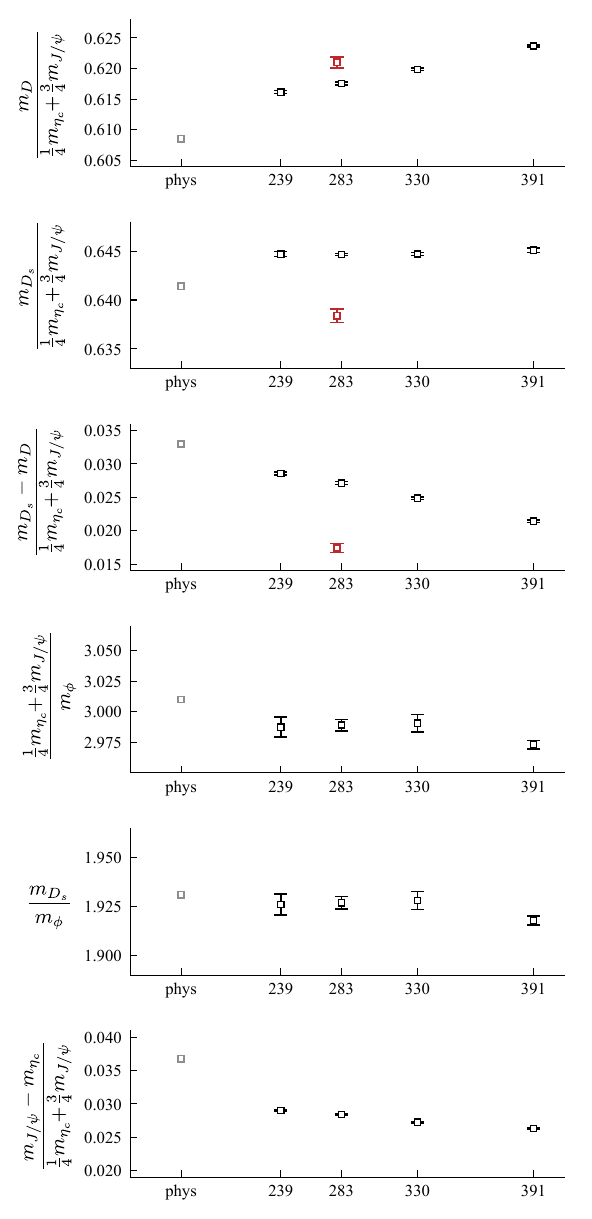}
  \caption{Ratios of stable hadron masses as a function of $m_\pi^2$. Black points determined in the current paper and Refs.\cite{Wilson:2023hzu,Wilson:2023anv}. Grey points taken from PDG values. Red points from Ref.~\cite{Prelovsek:2020eiw}. }
  \label{fig:mass_ratios}
\end{figure}

Discretization effects are not expected to be limited to deviations from the relativistic dispersion relation, and from a calculation at a single lattice spacing we cannot determine with confidence their form or magnitude. We do observe some systematic discrepancies which may originate in the lack of a continuum extrapolation, two of which are shown in Figure~\ref{fig:mass_ratios}. The bottom panel shows the `hyperfine splitting' scaled by the spin-averaged $1S$ charmonium mass as a function of light quark mass, which we observe to lie significantly below the experimental value, as has been observed previously~\cite{Liu:2012ze,Cheung:2016bym,Ryan:2020iog,Gayer:2024akw}. The top panel shows the $D$-meson mass ratio with the spin-averaged $1S$ charmonium mass, which we note appears to extrapolate to about 0.3\% above the physical value, indicating room for modest discretization effects at the few MeV level.

A reanalysis of the finite-volume spectrum of Ref.~\cite{Prelovsek:2020eiw} was presented in Ref.~\cite{Shi:2024llv} using amplitudes with improved analytic properties. While coming to quite similar conclusions about the pole content of the scattering system, this analysis had a great deal of difficulty describing both the rest-frame and moving-frame spectra simultaneously, having a $\chi^2/N_\mathrm{dof} > 3$ and visibly large departures from the computed spectrum at energies near the $D\bar{D}$ threshold. Potentially this could indicate that the finite-volume spectra are not compatible with description solely by scattering amplitudes in the finite-volume quantization condition -- this would be the case if there were non-negligible discretization effects.


\medskip
In Figure~\ref{fig:mass_ratios} we show evidence that in the current calculation, the charm quark is tuned consistently at each of the three pion masses considered, particularly in the fourth and fifth panels. In these panels we also see an indication that the charm mass may have been tuned very slightly low on the 391 MeV lattice reported in Refs.~\cite{Wilson:2023hzu,Wilson:2023anv}. The second panel illustrates the fact that the strange quark on the lattice used in Ref.~\cite{Prelovsek:2020eiw} is tuned to be lighter than the physical strange quark mass, which has the impact of significantly reducing the region between the $D\bar{D}$ threshold and the $D_s \bar{D}_s$ threshold (see the third panel).

The smaller of the two volumes used in Ref.~\cite{Prelovsek:2020eiw} has a spatial length in physical units of $L \sim 2.1\,\mathrm{fm}$ and ${m_\pi L \lesssim 3}$, and while this might be considered to be on the small side, pions probably do not play a primary role in the scattering processes of interest, and there are no obvious systematic discrepancies between the smaller volume and the larger, $L \sim 2.8\, \mathrm{fm}$, lattice. In the current calculation, the smaller of the 283 MeV lattices is of a similar size in physical units to the larger lattice in Ref.~\cite{Prelovsek:2020eiw}.

Ref.~\cite{Prelovsek:2020eiw} observes a difference in the $D$-meson mass determined on the two volumes, at the level of 0.2\%. As mentioned above, this variation is treated in an ad-hoc way with no corresponding uncertainty added. In the current calculation, on the $m_\pi \sim 283$ MeV lattices we observe a $D$-meson difference of a similar magnitude, and we choose to associate an additional $a_t \delta E = 0.00050$ systematic error on every energy level to conservatively account for it.



\begin{figure}
  \includegraphics[width=0.9\columnwidth]{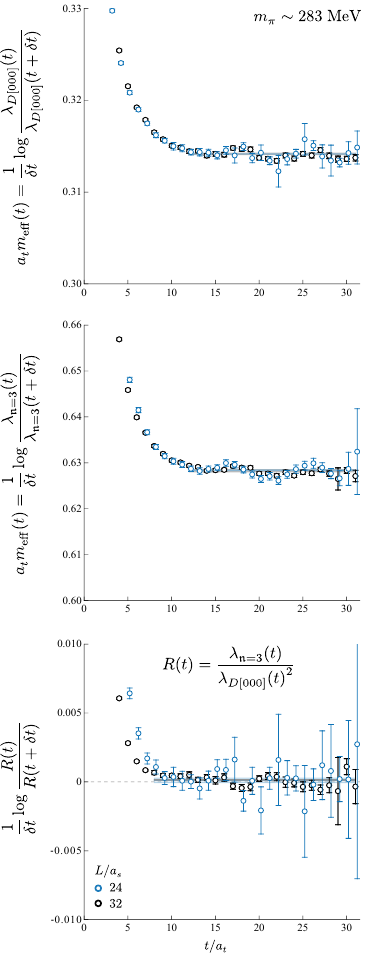}
  \caption{``Effective masses'' for the $D$-meson at-rest (top), the $D[000]\bar D[000]$-dominated level (middle), and a ratio of these (bottom). Each quantity is derived from the principal correlators $\lambda_\mathfrak{n}(t)$ determined from the variational method analysis of correlation matrices. A representative fit is shown in each case, where the width shows the fit region and the vertical extent shows the value and uncertainty.}
  \label{fig:corr_ratio}
\end{figure}


\begin{table}[b]
\centering
\renewcommand{\arraystretch}{1.2}
\begin{tabular}{lccc}
              & $m_\mathrm{eff}(t)$ & fit range & $\chi^2/N$ \\
\hline
\hline
$L/a_s = 24$  &              &            &      \\
$\lambda_{D[000]}(t)$   
      &  0.31415(9)  & $[12,31]$  & 0.87 \\
      &  0.31410(10) & $[15,31]$  & 0.91 \\
\hline
$\lambda_{\mathfrak{n}=3}(t)$ 
      &  0.62830(26) & $[12,31]$  & 1.50 \\
      &  0.62821(30) & $[15,31]$  & 1.50 \\
\hline
$R(t)$   
      &  0.00011(16) & $[8,31]$   & 0.57 \\
      & -0.00005(19) & $[12,31]$  & 0.39 \\
      & -0.00008(23) & $[15,31]$  & 0.46 \\
\hline
\hline
$L/a_s = 32$  &              &            &      \\
$\lambda_{D[000]}(t)$ 
      &  0.31415(6)  & $[12,31]$  & 2.2  \\
      &  0.31407(7)  & $[15,31]$  & 2.0  \\
\hline
$\lambda_{\mathfrak{n}=3}(t)$
      &  0.62839(11) & $[12,31]$  & 2.4  \\
      &  0.62820(13) & $[15,31]$  & 1.9  \\
\hline
$R(t)$   
      &  0.00015(3)  & $[8,31]$   & 2.4  \\
      &  0.00006(4)  & $[12,31]$  & 0.88 \\
      &  0.00001(5)  & $[15,31]$  & 0.90 \\
\hline
\hline
\end{tabular}
\caption{A selection of principal correlator and effective mass fits around $\DD$ threshold on the $m_\pi\sim283$~MeV lattices. $R(t)=\lambda_{\mathfrak{n}=3}/\lambda_{D[000]}^2$ as described in the text.}
\label{tab:corr_ratio_fits}
\end{table}


\section{Correlator ratios}
\label{app:corr_ratios}

Ref.~\cite{Prelovsek:2020eiw} determines their inputs for L\"uscher's method not by fitting correlation functions, but by forming a ratio of correlators and fitting that instead. From this, the energy difference between the non-interacting level and the level obtained from the correlation matrix (containing $\DD$ operators) can be determined. This assumes a perfect agreement with the continuum dispersion relation, even though there are differences that may reflect uncontrolled discretisation effects. In this appendix, we consider a ratio of principal correlators that mimics this approach somewhat, using the correlators obtained in this work.

For $[000]A_1^+$ on the $m_\pi\sim 283$~MeV ensembles, we consider the $\mathfrak{n}=3$ correlator, which in both cases is dominated by a $D[000]\bar{D}[000]$ operator construction and lies very close to the non-interacting energy from the variational analysis. Dividing this by the ground-state $[000]A_1^-$ $D[000]$ principal correlator squared gives the ratio $R(t)=\lambda_{\mathfrak{n}=3}(t)/\left(\lambda_{D[000]}(t)\right)^2$.  It has been observed in a few studies that excited state effects largely cancel in this ratio. Generally those studies use only ``simple'' meson-meson operators where the $D$ and $\bar{D}$-meson constituents are formed from a single operator ($\sim\bar{c}\gamma_5\ell$). In this work, projected operators are used to form the meson-meson operators and these are observed to reduce the excited-state contributions in the meson-meson correlators. Thus one does not expect a large effect at low $t$ values.

To illustrate this, we plot ``effective masses'' for these quantities, both individually and as a ratio, in Fig.~\ref{fig:corr_ratio} using 
\begin{equation}
a_t m_\mathrm{eff} = \frac{1}{\delta t}\mathrm{log}\frac{C(t)}{C(t+\delta t)}
\label{eq:effmass}
\end{equation}
where we have chosen $\delta t/a_t=3$. We work from the principal correlators $\lambda_{\mathfrak{n}}$ determined from the variational method in the main analysis. We summarise a few fits of these using simple constants to the large time behaviour in Table~\ref{tab:corr_ratio_fits}. A few different fit ranges are considered. In both cases, the numerators ($\lambda_{\mathfrak{n}=3}$) match exactly those used in the main analysis above. We form the ratio under jackknife, averaged over the same time sources for numerator and denominator in order to allow any correlated fluctuations to cancel as much as possible. At large $t$ values, we observe every quantity is dominated by a single exponential and thus tends to a constant in the effective mass equation~\ref{eq:effmass}. For the ratio $R(t)$, this should correspond to the difference $\Delta E = E_{\mathfrak{n}=3} - 2E_{D[000]}$.

We see that in this approach (where the systematic differences in $D$-meson masses are neglected) we do find smaller statistical uncertainties than reported in the main study. However, the differences $\Delta E = E_{\mathfrak{n}=3} - 2E_{D[000]}$ are still very close to (or consistent with) zero, pointing to insignificant interactions at threshold.

\end{document}